\documentclass[prb]{revtex4}

\usepackage{amsmath}
\usepackage{graphicx}
\usepackage{bm}

\newcommand{\bra}[1]{\langle #1|}
\newcommand{\ket}[1]{|#1\rangle}

\begin{document}

\title{Lagrangian approach to the semi-relativistic electron dynamics in the mean-field approximation}
\author{Anant Dixit, Yannick Hinschberger, Jens Zamanian, Giovanni Manfredi\footnote{Corresponding author: giovanni.manfredi@ipcms.unistra.fr}, Paul-Antoine Hervieux}
\affiliation{Institut de
Physique et Chimie des Mat\'{e}riaux, CNRS and Universit\'{e} de
Strasbourg, BP 43, F-67034 Strasbourg, France}

\date{\today}

\begin{abstract}
We derive a mean-field model that is based on a two-component Pauli-like equation and incorporates quantum, spin, and relativistic effects up to second order in $1/c$. Using a Lagrangian approach, we obtain the self-consistent charge and current densities that act as sources in the Maxwell equations. A physical interpretation is provided for the second-order corrections to the sources. The Maxwell equations are also expanded to the same order. The resulting self-consistent model constitutes a suitable semi-relativistic approximation to the full Dirac-Maxwell equations.
\end{abstract}


\maketitle

\section{Introduction}

The interaction of a femtosecond electromagnetic pulse with the electron spin in a ferromagnetic metal has been the object of intense investigations, both theoretical and experimental, during the past fifteen years \cite{Bigot_annphys}.
Typical experiments involve
perturbing the electron charge and spin with an ultrashort light pulse (the pump), followed by a second weaker pulse (the probe) that acts as a diagnostic tool. By modulating
the relative amplitude of the signals, as well as the delay
between the pump and the probe, it is possible to assess with
great precision the dynamical relaxation of the electron gas.

Early results \cite{Beaurep} already pointed at the quick loss of magnetization that occurs following the excitation by a femtosecond laser pulse. These experiments showed that the electron spins respond to the excitation on a subpicosecond timescale, which is a typical timescale for electrons to equilibrate thermally with the lattice in metal nanostructures. These thermal processes can be explained qualitatively by ``three-temperature models", involving the temperatures of the electron, the lattice and the spin, without specifying the exact nature of the interaction between the spins and the charges. From a fundamental point of view, several mechanisms have been proposed for the modification of the magnetic order of nanostructures subject to an ultrafast external field, ranging from the spin-orbit coupling \cite{Zhang} to the spin-lattice interactions \cite{Koopmans}. A recent review of the state of the art in the field of ultrafast magnetization dynamics in nanostructures can be found in Ref. \cite{Bigot_annphys}.

Recent experiments \cite{Bigot_natphys} have now given a new twist to these problems, with promising future developments, both theoretical and experimental. These experiments have shown the existence of a coherent coupling between a femtosecond laser pulse and the magnetization of a ferromagnetic thin film. The underlying mechanism is thought to involve a form of spin-orbit coupling (SOC) that goes beyond the usual one due to the electric field of the ions, and involves the electromagnetic field of the laser pulse \cite{Vonesch}. This coherent mechanism is clearly distinguished from the incoherent ultrafast demagnetization associated with the thermalization of the spins.
The theoretical description of these coherent effects is still lacking and will be mandatory in order to gain a sound understanding of ultrafast laser-spin interactions \cite{Hinsch_prb}.

The electromagnetic field associated with a femtosecond laser pulse can be strong enough to significantly perturb the electronic charges and spins in nanometric systems, so that relativistic effects become important. Given the intensity of the fields involved, nonlinear effects are also expected to play a considerable role. This scenario represents an ambitious theoretical challenge, as it requires the modelling of the nonlinear dynamics of a quantum-relativistic system of many interacting electrons excited by an intense and ultrashort electromagnetic field.

In neighboring areas, relativistic corrections to the many-electron dynamics are taken into account in quantum chemistry calculations, particularly for heavy elements \cite{Keller}. Relativistic versions of density functional theory (DFT) based on the Dirac-Kohn-Sham equations and relativistic mean-field or Dirac-Hartree-Fock models have been developed for these purposes  \cite{Dyall,Engel, Rajagopal, Parpia, Romaniello}, but they are in general rather complex to handle either analytically or numerically.

The goal of the present work is to develop a time-dependent semi-relativistic mean-field theory that is based on two-component wave functions and thus considerably simpler than the full relativistic models relying on the Dirac equation. The model should preserve the mathematical structure of the Schr\"{o}dinger or Kohn-Sham equations, which have been implemented in many numerical codes with great computational sophistication.

Semi-relativistic approximation to the Dirac equation can be formally obtained in several ways, for instance by making use of the Foldy-Wouthuysen transformation \cite{Foldy,Hinsch} to expand the Dirac equation in powers of the inverse of the speed of light in vacuum $c$. To lowest order, the only correction to the Schr\"{o}dinger equation comes from the Zeeman term, coupling the spin to the magnetic field. Nevertheless, it is well known that second-order effects such as the SOC are crucial for the proper understanding of magneto-optical processes and should therefore be retained. Hence the need to develop a self-consistent second-order approach.

It has been noted that the Hamiltonian obtained through Foldy-Wouthuysen transformation is not a regular approximation of the full Dirac Hamiltonian, because the expansion of the relativistic kinetic energy has a finite radius of convergence. Thus, the expansion is no longer valid for large enough values of the momentum $\mathbf{p}$, which is bound to occur as the Coulomb potential of the nuclei diverges as $1/r$. Quantum chemistry calculations generally resort to the so-called ZORA (zeroth-order regular approximation) equation in order to circumvent this problem \cite{Van_Lenthe}.
In the present work, we shall neglect altogether the relativistic correction to the kinetic energy, because it introduces fourth-order spatial derivatives that are difficult to conciliate with a Schr\"{o}dinger-like equation (The effect of such a term on the current density is however described in Appendix \ref{App:Rel_Corr}).

Things become even more complicated when the Dirac equation is coupled self-consistently to the Maxwell equations for the electromagnetic fields, for the resulting system is nonlinear, in contrast to the linear Dirac or Schr\"{o}dinger equations. In addition, for the internal consistency of the overall model, one should also require that the Maxwell equations be expanded to the same order in $1/c$ \cite{Holland-Brown,Sulaksono}.

An important issue lies in the determination of the sources to be inserted into the Maxwell equations. In a fully relativistic approach, the Dirac 4-current is the appropriate expression; conversely, for the nonrelativistic Schr\"{o}dinger or Pauli equations, the density and current are also well-established. For semi-relativistic models to a certain order in $1/c$ the situation is more complicated and no clear-cut consensus exists on this issue.

In the present work, the Dirac equation is approximated to second order in $1/c$ to obtain a Pauli-type Hamiltonian for the two-component wave function (hereafter, we refer to this Hamiltonian as the ``extended Pauli Hamiltonian" and the corresponding equation as the `extended Pauli equation", or EPE).
Next, we derive a Lagrangian density that reproduces such EPE as well as the Maxwell equations.

By using the standard minimal coupling between the sources and the electromagnetic fields, we are able to obtain the relativistic corrections to the classical probability density and the probability current density.
We then derive the expression for charge and mass conservation from different methods (direct calculation, Noether's theorem) thereby verifying the correctness of the sources obtained with the Lagrangian approach.
A physical interpretation of the various correction terms is attempted whenever possible.

Finally, the Maxwell equations are also expanded to second order to maintain consistency with the second-order EPE \cite{manfredi_maxwell}. The resulting model contains all the standard semi-relativistic terms (Zeeman effect, Darwin term, spin-orbit coupling, Hartree mean field) in addition to the self-consistent magnetic fields generated by the internal currents. The model
should constitute an appropriate self-consistent approximation of the Dirac-Maxwell equations to second order in $1/c$.

\section{Lagrangian formalism -- General framework}
We consider a many-electron system where both quantum and relativistic effects can in principle play a significant role. In a mean-field approach, the electron dynamics is governed by the Dirac equations
\begin{equation}\label{Eq:Dirac_Hart}
i\hbar~ \frac{\partial \Psi^n_{D}}{\partial t}=c\boldsymbol{\alpha}\cdot(\hat{\mathbf{p}}- q\mathbf{A})\Psi^n_{D}+\beta mc^2\Psi^n_{D}+q\phi\Psi^n_{D}
\end{equation}
coupled self-consistently to the Maxwell equations, which we write in terms of the scalar and vector potentials $(\phi$, $\mathbf{A})$ in the Lorentz gauge
($\nabla\cdot \mathbf{A} + c^{-2}\partial_t \phi = 0 $):
\begin{eqnarray}
-\Delta \phi + \frac{1}{c^2}\frac{\partial^2 \phi}{\partial t^2} &=& \frac{q\rho}{\varepsilon_0}, \label{Eq:max_phi}\\
-\Delta \mathbf{A} + \frac{1}{c^2}\frac{\partial^2 \mathbf{A}}{\partial t^2} &=& \mu_0 q\mathbf{j}~, \label{Eq:max_A}
\end{eqnarray}
and the sources are given by the Dirac 4-current, i.e.:
\begin{equation}
(c\rho,\mathbf{j}) = c\sum_n(\Psi^{n \dag}_{D}\Psi^n_{D}, \Psi^{n \dag}_{D}\boldsymbol{\alpha} \Psi^n_{D}).
\end{equation}

Here, $\Psi^n_{D}$ represents the Dirac bispinor for the $n$th electron,
$\boldsymbol{\alpha}$ and $\beta$ are the usual Dirac matrices, $m$ and $q$ are the particle rest mass and charge (for electrons $m=m_e$ and $q=-e$), $\varepsilon_0$ and $\mu_0$ are the electric permittivity and the magnetic permeability in vacuum ($\varepsilon_0 \mu_0 c^2= 1$), $\hbar$ is the Planck constant, $\rho$ is the probability density, and ${\mathbf j}$ is the probability current density.

Equations (\ref{Eq:Dirac_Hart})--(\ref{Eq:max_A}) constitute a fully relativistic, Lorentz-covariant model for the quantum dynamics of a system of $N$ electrons in the mean-field approximation.

The purpose of the present paper is to work out a semi-relativistic version of this model that is valid to second order in $1/c$. This low-energy semi-relativistic theory should consider only electrons and neglect positrons (negative energy states) and thus should be based on two-component spinors. Ignoring issues of self-consistency for the time being, the second-order Hamiltonian can be obtained from the full Dirac Hamiltonian by performing a Foldy-Wouthuysen transformation \cite{Foldy,Hinsch}. One obtains:
\begin{equation}
\hat{H}=mc^2+q\phi+\frac{(\hat{\mathbf{p}}-q\mathbf{A})^2}{2m}-
\frac{q\hbar}{2m}\boldsymbol{\sigma}\cdot\mathbf{B}- \frac{(\hat{\mathbf{p}}-q\mathbf{A})^4}{8m^3c^2}
-\frac{q\hbar^2}{8m^2c^2} \nabla\cdot\mathbf{E}-\frac{q\hbar}{8m^2c^2}\boldsymbol{\sigma} \cdot[\mathbf{E}\times(\hat{\mathbf{p}}-q\mathbf{A})+(\hat{\mathbf{p}} -q\mathbf{A})\times\mathbf{E}]
\label{Eq:hamiltonian}
\end{equation}
where the electromagnetic fields are defined as usual as: $\mathbf{E}=-\nabla\phi-\partial_t\mathbf{A}$ and $\mathbf{B}=\nabla\times\mathbf{A}$.

Here, the first term on the right-hand side is the rest-mass energy of the electron; the next two terms are the standard Schr\"{o}dinger Hamiltonian in the presence of an electromagnetic field; the fourth term is the Pauli spin term (Zeeman effect); the $(\hat{\mathbf{p}}-q\mathbf{A})^4$ term is the first relativistic correction to the electron mass (expansion of the Lorentz factor $\gamma$ to second order); the $\nabla \cdot \mathbf{E}$ term is the Darwin term; and the last two terms represent the spin-orbit coupling (SOC).

In terms of the vector and scalar potentials, the Hamiltonian (\ref{Eq:hamiltonian}) can also be written in the following form:
\begin{equation}\label{Eq:hamiltonian_pot}
\begin{aligned}
\hat{H}&=mc^2+q\phi+\frac{(\hat{\mathbf{p}}-q\mathbf{A})^2}{2m}- \frac{q\hbar}{2m}\boldsymbol{\sigma}\cdot(\nabla\times\mathbf{A})+ \frac{q\hbar^2}{8m^2c^2}\Delta\phi+ \frac{q\hbar^2}{8m^2c^2}\nabla\cdot\partial_t\mathbf{A}\\
&-\frac{q\hbar}{4m^2c^2}\boldsymbol{\sigma}\cdot\left[(\nabla \phi+\partial_t\mathbf{A})\times(\hat{\mathbf{p}}-q\mathbf{A}) \right]-\frac{q\hbar}{8m^2c^2}\boldsymbol{\sigma}\cdot(\hat{\mathbf{p}} \times\partial_t\mathbf{A}).
\end{aligned}
\end{equation}

From now on, we will neglect the relativistic correction to the electron mass [i.e., the term proportional to $(\hat{\mathbf{p}}-q\mathbf{A})^4$ in the expansion of the kinetic energy]. Although it is still second-order in $1/c$, this term introduces fourth-order derivatives in the evolution equation, unlike the non-relativistic Schr\"{o}dinger equation which only contains second-order derivatives. A Lagrangian density that accounts for the relativistic mass correction can nevertheless be found, as is shown in Appendix \ref{App:Rel_Corr}, where we also provide the resulting corrections on the current density (the particle density is unchanged).

The resulting extended Pauli equation
\begin{equation}\label{Eq:Ext_Pauli}
i\hbar \frac{\partial \Psi}{\partial t} = \hat{H} \Psi,
\end{equation}
can be derived from a Lagrangian density by applying the Euler-Lagrange equations. One form of the Lagrangian density is
\begin{equation}\label{Eq:Lagr_Anant}
\mathcal{L}=\frac{1}{2}\Psi^\dagger\left[(i\hbar\partial_t-\hat{H}) \Psi\right]+\frac{1}{2}\left[(-i\hbar\partial_t-\hat{H}^\dagger)\Psi^\dagger\right]\Psi,
\end{equation}
but the following form also yields the same Pauli equation:
\begin{equation}\label{Eq:Lagr_Jens}
\mathcal{L}=\Psi^\dagger(i\hbar\partial_t-\hat{H})\Psi.
\end{equation}

The Lagrangian density we propose is based on a combination of the above forms and is defined as:
\begin{equation}\label{Eq:Lagr_Dens}
\begin{aligned}
\mathcal{L}_P=&\frac{i\hbar}{2}(\Psi^\dagger\dot{\Psi}-\dot{\Psi^\dagger}\Psi)-\Psi^\dagger (mc^2+q\phi)\Psi-\frac{1}{2m}[(i\hbar\partial_k-q A_k)\Psi^\dagger(-i\hbar\partial_k-q A_k)\Psi]\\
&+\Psi^\dagger\left[\frac{q\hbar}{2m}\epsilon_{ijk}\sigma_i\partial_jA_k-\frac{q\hbar^2}
{8m^2c^2}\partial_k^2\phi-\frac{q\hbar^2}{8m^2c^2}\partial_t\partial_k A_k\right]\Psi\\
&+\Psi^\dagger\epsilon_{ijk}\left[\frac{q\hbar}{4m^2c^2}\sigma_i \partial_j\phi q A_k+\frac{q\hbar}{4m^2c^2}\sigma_i \partial_tA_j q A_k\right]\Psi\\
&-\frac{q\hbar}{8m^2c^2}\epsilon_{ijk}[\Psi^\dagger\sigma_i\partial_j\phi\hat{p}_k\Psi-
\partial_j\phi\hat{p}_k\Psi^\dagger\sigma_i \Psi +\Psi^\dagger\sigma_i\partial_t A_j\hat{p}_k\Psi-\partial_tA_j \hat{p}_k \Psi^\dagger\sigma_i\Psi],
\end{aligned}
\end{equation}
where $\epsilon_{ijk}$ is the Levi-Civita symbol (equal to $+1$ for even permutations of $ijk$, $-1$ for odd permutations, and $0$ if any index is repeated) and we have assumed the Einstein summation convention over repeated indexes.
Also note that from now on, we shall omit the summation $\sum_n$ over all particles of the system, for simplicity of notation.

The advantage of the Lagrangian approach is that the self-consistency can be incorporated in the model simply by adding the Lagrangian for the electromagnetic fields. In the Lorentz gauge, this reads as:
\begin{equation}
\mathcal{L}_{EM} = \frac{\varepsilon_0}{2}(\partial_k\phi)^2-\frac{\varepsilon_0}{2c^2}(\partial_t\phi)^2- \frac{1}{2\mu_0}(\partial_jA_k)^2+\frac{1}{2\mu_0 c^2}(\partial_t A_k)^2.
\end{equation}

The total Lagrangian (particles and fields) is then $\mathcal{L}(\Psi,\Psi^\dagger,\phi,\mathbf{A}) = \mathcal{L}_{P}+\mathcal{L}_{EM}$. By taking the Euler-Lagrangian equation for $\mathcal{L}$ with respect to the scalar and vector potentials, we obtain the Maxwell equations (\ref{Eq:max_phi})-(\ref{Eq:max_A}), with certain (yet undetermined) expressions on the right-hand side of the equations. These expressions can be identified as the particle and current densities of our system, governed by the EPE.
Thus, we have a systematic and straightforward method to obtain the second-order relativistic corrections to the standard (Schr\"{o}dinger) expressions of the density and current.

The Euler-Lagrange equation for a Lagrangian density depending on up to the second derivatives of a field $\varphi$ reads as \cite{Durr}:
\begin{equation}\label{Eq:E-L_Eq}
\frac{\partial\mathcal{L}}{\partial\varphi}- \sum_\mu\partial_\mu\frac{\partial\mathcal{L}}{\partial(\partial_\mu\varphi)}+ \sum_\mu\partial_\mu^2\frac{\partial\mathcal{L}}{\partial(\partial_\mu^2\varphi)}+ \sum_{\mu\neq\nu}\partial_\mu\partial_\nu\frac{\partial\mathcal{L}} {\partial(\partial_\mu\partial_\nu\varphi)}=0,
\end{equation}
where the subscripts $\mu,~\nu$ denote the space-time coordinates $(ct,x,y,z)$ and the field $\varphi$ is either the scalar potential or one Cartesian component of the vector potential. The first two terms in Eq. (\ref{Eq:E-L_Eq}) are the standard ones for a Lagrangian that depends only on the first derivatives.
We need this general form of the Euler-Lagrange equations for our system since the terms we encounter contain mixed derivatives up to second order.

The Lagrangian density (\ref{Eq:Lagr_Dens}) does return the EPE when we calculate the Euler-Lagrange equations for $\Psi^\dagger$, as is shown in Appendix \ref{App:Lagr_Test}.

\section{Relativistic corrections to the sources}
Our goal here is to compute the relativistic corrections to the sources (density and current) that appear in the Maxwell equations.
In principle, one could proceed straight from the Dirac current by applying the Foldy-Wouthuysen transformations on the Dirac 4-current, but the procedure would be mathematically complicated.
Here, we will show that the correct results can be obtained by means of the Lagrangian approach outlined in the preceding section.

As we shall see, the obtained charge and current densities contain several terms beyond the result of standard relativistic DFT \cite{Romaniello}.

\subsection{Probability density}
We compute each term appearing in the Euler-Lagrange equations for $\phi$:
\begin{equation}
\frac{\partial\mathcal{L}}{\partial\phi}=- q\Psi^\dagger\Psi \nonumber
\end{equation}
\begin{equation}
\partial_t\frac{\partial\mathcal{L}}{\partial(\partial_t\phi)}= -\frac{\varepsilon_0}{c^2} \frac{\partial^2 \phi}{\partial t^2} \nonumber
\end{equation}
\begin{equation}
\frac{\partial\mathcal{L}}{\partial(\partial_j\phi)}=\frac{q\hbar}{4m^2c^2} [q\mathbf{A}\times(\Psi^\dagger\boldsymbol{\sigma}\Psi)]_j-\frac{iq\hbar^2}{8m^2c^2} [(\nabla\Psi^\dagger)\times\boldsymbol{\sigma}\Psi+\Psi^\dagger \boldsymbol{\sigma}\times(\nabla\Psi)]_j+\varepsilon_0\partial_j\phi \nonumber
\end{equation}
\begin{equation}
\partial_j^2\frac{\partial\mathcal{L}}{\partial(\partial_j^2\phi)}= -\frac{q\hbar^2}{8m^2c^2}\Delta(\Psi^\dagger\Psi). \nonumber
\end{equation}

Combining the above terms in the Euler-Lagrange Eq. (\ref{Eq:E-L_Eq}) for $\phi$, we obtain
\begin{equation}
\begin{aligned}
-\varepsilon_0\Delta\phi+\frac{\varepsilon_0}{c^2} \frac{\partial^2 \phi}{\partial t^2}&=q\Psi^\dagger\Psi+\frac{q\hbar^2}{8m^2c^2}\Delta(\Psi^\dagger\Psi)\\
&+\frac{q\hbar}{4mc^2}\nabla\cdot\left[\frac{q}{m}\mathbf{A} \times(\Psi^\dagger\boldsymbol{\sigma}\Psi)- \frac{i\hbar}{2m}\{(\nabla\Psi^\dagger)\times\boldsymbol{\sigma}\Psi+ \Psi^\dagger\boldsymbol{\sigma}\times(\nabla\Psi)\}\right].
\end{aligned}
\end{equation}

This allows us, using Eq. (\ref{Eq:max_phi}), to define the probability density as:
\begin{equation}\label{Eq:Prob_dens}
\rho\equiv\Psi^\dagger\Psi+ \frac{\hbar}{4mc^2}\nabla\cdot\left\{\frac{\hbar}{2m} \nabla(\Psi^\dagger\Psi)\right\}+ \frac{\hbar}{4mc^2}\nabla\cdot\left[\frac{q}{m}\mathbf{A} \times(\Psi^\dagger\boldsymbol{\sigma}\Psi)- \frac{i\hbar}{2m}\{(\nabla\Psi^\dagger)\times\boldsymbol{\sigma}\Psi+ \Psi^\dagger\boldsymbol{\sigma}\times(\nabla\Psi)\}\right].
\end{equation}

At zeroth order, we recover the standard Schr\"{o}dinger or Pauli density $\Psi^\dagger\Psi$. Further, we see that all correction terms at second order in $1/c$ can be written as the divergence of a vector $\mathbf{P}$, which can be interpreted as a polarization density.

\subsection{Probability current density}
In order to obtain the expression of the current density, we compute the Euler-Lagrange equation (\ref{Eq:E-L_Eq}) for the $k$-th component of the vector potential $A_k$. Explicitly, this yields:
\begin{equation}
\frac{\partial\mathcal{L}}{\partial A_k}=\frac{iq\hbar}{2m}[(\nabla\Psi^\dagger)\Psi-\Psi^\dagger(\nabla\Psi)]_k- \frac{q^2}{m}A_k\Psi^\dagger\Psi-\frac{q^2\hbar}{4m^2c^2} [(\nabla\phi+\partial_t\mathbf{A})\times(\Psi^\dagger\boldsymbol{\sigma}\Psi)]_k.
\nonumber
\end{equation}
\begin{equation}
\frac{\partial\mathcal{L}}{\partial(\partial_t A_j)}=\frac{q^2\hbar}{4m^2c^2}[\mathbf{A}\times(\Psi^\dagger\boldsymbol{\sigma}\Psi)]_j- \frac{qi\hbar^2}{8m^2c^2}\frac{\partial}{\partial t}[(\nabla\Psi^\dagger)\times\boldsymbol{\sigma}\Psi+ \Psi^\dagger\boldsymbol{\sigma}\times(\nabla\Psi)]_j+\frac{1}{\mu_0 c^2}\partial_t A_j . \nonumber
\end{equation}
\begin{equation}
\frac{\partial\mathcal{L}}{\partial(\partial_j A_k)}=\frac{q\hbar}{2m}\epsilon_{ijk}\Psi^\dagger\boldsymbol{\sigma}_i\Psi- \frac{1}{\mu_0}\partial_jA_k.
\nonumber
\end{equation}
\begin{equation}
\frac{\partial\mathcal{L}}{\partial(\partial_t \partial_j A_k)}=-\frac{q\hbar^2}{8m^2c^2}\delta_{jk}\Psi^\dagger\Psi.
\nonumber
\end{equation}

Combining the above terms together gives the Maxwell equation for $\mathbf{A}$:
\begin{equation}
\begin{aligned}
\frac{1}{\mu_0 c^2}\frac{\partial^2\mathbf{A}}{\partial t^2}-\frac{1}{\mu_0}\Delta
\mathbf{A}=&\frac{iq\hbar}{2m}[(\nabla\Psi^\dagger) \Psi-\Psi^\dagger(\nabla\Psi)]-\frac{q^2}{m} \mathbf{A}\Psi^\dagger\Psi-\frac{q^2\hbar}{4m^2c^2}\partial_t[\mathbf{A} \times(\Psi^\dagger\boldsymbol{\sigma}\Psi)]\\
&-\frac{q^2\hbar}{4m^2c^2}[(\nabla\phi+\partial_t \mathbf{A})\times(\Psi^\dagger\boldsymbol{\sigma}\Psi)]+ \frac{q\hbar}{2m}[\nabla\times(\Psi^\dagger\boldsymbol{\sigma}\Psi)]\\
&+\frac{qi\hbar^2}{8m^2c^2}[(\nabla\Psi^\dagger) \times\boldsymbol{\sigma}\Psi+\Psi^\dagger\boldsymbol{\sigma} \times(\nabla\Psi)]-\frac{q\hbar^2}{8m^2c^2} \partial_t[\nabla(\Psi^\dagger\Psi)].
\end{aligned}
\end{equation}
Comparing with Eq. (\ref{Eq:max_A}), we find that the probability current density can be defined as:
\begin{equation}\label{Eq:Prob_Curr}
\begin{aligned}
\mathbf{j}\equiv &\frac{i\hbar}{2m}[(\nabla\Psi^\dagger)\Psi- \Psi^\dagger(\nabla\Psi)]-\frac{q}{m}\mathbf{A}\Psi^\dagger\Psi+ \frac{\hbar}{2m}[\nabla\times(\Psi^\dagger\boldsymbol{\sigma}\Psi)]+ \frac{q\hbar}{4m^2c^2}[\mathbf{E}\times(\Psi^\dagger\boldsymbol{\sigma}\Psi)]\\
-&\frac{\hbar}{4mc^2}\frac{\partial}{\partial t}\left[\frac{\hbar}{2m}\nabla (\Psi^\dagger\Psi)\right]-\frac{\hbar}{4mc^2}\frac{\partial}{\partial t}\left[\frac{q}{m}\mathbf{A} \times(\Psi^\dagger\boldsymbol{\sigma}\Psi)- \frac{i\hbar}{2m}[(\nabla\Psi^\dagger)\times\boldsymbol{\sigma}\Psi+ \Psi^\dagger\boldsymbol{\sigma}\times(\nabla\Psi)]\right].
\end{aligned}
\end{equation}

At zeroth order, we recover the Schr\"{o}dinger current density along with the usual spin current term $\frac{\hbar}{2m}\nabla\times(\Psi^\dagger\boldsymbol{\sigma}\Psi)$ \cite{Landau}, which is the curl of a magnetization vector and therefore does not appear in the continuity equation. An interpretation of the current and density corrections at second order is provided in the forthcoming section.

\section{Discussion and Interpretation of the Sources}
The charge and current densities can be rewritten as the sum of a free part and a bound part:
\begin{eqnarray}
q\rho&=&q\rho^{free}-\nabla\cdot\mathbf{P},\\
q\mathbf{j}&=&q\mathbf{j}^{free}+\nabla\times\mathbf{M}+ \partial_t\mathbf{P},
\end{eqnarray}
where
\begin{eqnarray}
\rho^{free}&=&\Psi^\dagger\Psi, \label{Eq:rho_free}\\
\mathbf{j}^{free}&=&\frac{i\hbar}{2m}[(\nabla\Psi^\dagger)\Psi- \Psi^\dagger(\nabla\Psi)]-\frac{q}{m}\mathbf{A}\Psi^\dagger\Psi+ \frac{q\hbar}{4m^2c^2}[\mathbf{E}\times(\Psi^\dagger\boldsymbol{\sigma}\Psi)], \label{Eq:j_free}\\
\mathbf{M}&=&\mathbf{M}_{spin}=\frac{qh}{2m}(\Psi^\dagger\boldsymbol{\sigma}\Psi),\\
\mathbf{P}_{spin}&=&-\frac{q\hbar}{4mc^2}\left[\frac{q}{m} \mathbf{A}\times(\Psi^\dagger\boldsymbol{\sigma}\Psi)-\frac{i\hbar}{2m} \{(\nabla\Psi^\dagger)\times\boldsymbol{\sigma}\Psi+\Psi^\dagger\boldsymbol{\sigma}\times(\nabla\Psi)\}\right], \label{Eq:spin_pol}\\
\mathbf{P}_{Darwin}&=&-\frac{q\hbar^2}{8m^2c^2}\nabla(\Psi^\dagger\Psi).
\label{Eq:darwin_pol}
\end{eqnarray}
The polarization density $\mathbf{P}=\mathbf{P}_{spin}+\mathbf{P}_{Darwin}$ has been written as the sum of a ``spin" polarization and a ``Darwin" polarization, hinting at the origin of these two terms.
We also note that the free density does not contain any second-order corrections. Instead, the free current density displays a correction term that can be written as $(\mathbf{E} \times \mathbf{M}_{spin})/2mc^2$. This term was already obtained in the past from semi-relativistic kinetic models (Wigner equation) \cite{Zamanian}.

We will now attempt to provide a physical interpretation of these terms.
Let us start with the spin polarization.
We consider two reference frames moving with a velocity $\mathbf{v}$ with respect to each other. The Lorentz transformations for the magnetization and  polarization vectors read as \cite{Strange}
\begin{eqnarray}
\mathbf{P}&=&\gamma\left(\mathbf{P}^\prime+\frac{\mathbf{v}\times\mathbf{M}^\prime} {c^2}\right)-\frac{\gamma^2}{1+\gamma}\left(\frac{\mathbf{P}^\prime\cdot\mathbf{v}} {c}\right)\frac{\mathbf{v}}{c}\\
\mathbf{M}&=&\gamma\left(\mathbf{M}^\prime-\mathbf{v}\times\mathbf{P}^\prime\right) -\frac{\gamma^2}{1+\gamma}\left(\frac{\mathbf{M}^\prime\cdot\mathbf{v}}{c}\right) \frac{\mathbf{v}}{c},
\end{eqnarray}
where $\gamma=(1-v^2/c^2)^{-1/2}$ is the usual Lorentz factor.
For $v\ll c$ and $|\mathbf{M}| \gg c|\mathbf{P}|$ (electric limit) the above transformations become
\begin{eqnarray}
\mathbf{P}&=&\mathbf{P}^\prime+\frac{\mathbf{v}\times\mathbf{M}^\prime}{c^2} \label{Eq:polarization_lor}\\
\mathbf{M}&=&\mathbf{M}^\prime.
\label{Eq:magnetization_lor}
\end{eqnarray}
In the rest frame of the electron (primed variables) there is a magnetization  $\mathbf{M}' = \mathbf{M}_{spin}$, but no polarization, $\mathbf{P}'=0$. Thus, in the laboratory frame (unprimed variables), we have
\begin{eqnarray}
\mathbf{P}&=&\frac{\mathbf{v}\times\mathbf{M}_{spin}}{c^2} \\
\mathbf{M}&=&\mathbf{M}_{spin}.
\end{eqnarray}

The above line of reasoning is purely classical, in the sense that $\mathbf{v}$ and $\mathbf{P}$ are real numbers, not operators. To compare with the quantum result (\ref{Eq:spin_pol}), we define the velocity operator as: $\hat{\mathbf{v}}=(\hat{\mathbf{p}}-q\mathbf{A})/m$, and the magnetization operator as
\begin{equation}
\hat{\mathbf{M}}=\frac{qh}{2m}~\boldsymbol{\sigma}.
\end{equation}
Then we can define a polarization operator:
\begin{equation}
\hat{\mathbf{P}}_{spin}= \frac{\hat{\mathbf{v}}\times\hat{\mathbf{M}}_{spin}}{c^2}
= -\frac{q\hbar}{2m^2c^2}\left(q\mathbf{A} \times \boldsymbol{\sigma}-\frac{1}{2}
\hat{\mathbf{p}} \times \boldsymbol{\sigma} +
\frac{1}{2}
\boldsymbol{\sigma} \times \hat{\mathbf{p}} \right).
\end{equation}
Making use of $\hat{\mathbf{p}}=-i\hbar\nabla$, we obtain the expectation value of $\hat{\mathbf{P}}_{spin}$
\begin{equation}
\langle \Psi|\hat{\mathbf{P}}_{spin}|\Psi \rangle= -\frac{q\hbar}{2m^2c^2}\left[q\mathbf{A} \times \langle \Psi|\boldsymbol{\sigma}|\Psi\rangle +
\frac{i\hbar}{2}\Big(\langle\Psi|\nabla\times\boldsymbol{\sigma}|\Psi\rangle-
\langle\Psi|\boldsymbol{\sigma}\times\nabla|\Psi\rangle\Big)
\right], \label{Eq:polarization}
\end{equation}
which, upon an integration by parts, becomes identical to the spatial average of the spin polirization vector as defined in Eq. (\ref{Eq:spin_pol}), except for a factor 2. This factor has the same origin as the well-known Thomas correction in the spin-orbit Hamiltonian.
Thus, we have seen that the spin polarization is a manifestation of the spin magnetization in the laboratory frame of reference.

Let us now turn to the Darwin polarization. It is usually admitted that the Darwin term in the Hamiltonian is a manifestation of the so-called \emph{Zitterbewegung}, i.e., a quivering motion of the electron around its mean path \cite{Sakurai}. This is due to the interference between the positive and negative energy states, which produce fluctuations of the position of the particle. In general, this is used to find the Darwin correction to the Hamiltonian, but here we show that the Zitterbewegung is also at the origin of the second-order density correction.

To understand how the Darwin term is a manifestation of this phenomenon, we consider fluctuations $\delta{\mathbf{r}}(t)$ around the mean trajectory $\overline{{\mathbf{r}}(t)}$:
\begin{equation}
{\mathbf{r}}(t)=\overline{{\mathbf{r}}(t)}+\delta{\mathbf{r}}(t).
\end{equation}

We expand the probability distribution of the particle around the mean position. We obtain (neglecting the time dependence and using Einstein's summation convention for simplicity of notation):
\begin{equation}
\rho({\mathbf{r}})=\Psi^\dagger(\overline{{\mathbf{r}}}) \Psi(\overline{{\mathbf{r}}})+\nabla[\Psi^\dagger(\overline{{\mathbf{r}}}) \Psi(\overline{{\mathbf{r}}})]\cdot\delta{\mathbf{r}}+\frac{1}{2} \frac{\partial^2[\Psi^\dagger(\overline{{\mathbf{r}}}) \Psi(\overline{{\mathbf{r}}})]}{\partial r_i \partial r_j}~\delta r_i\delta r_j+\cdots
\end{equation}

The perturbation due to the \emph{Zitterbewegung} is $\rho_Z\equiv \overline{\rho({\mathbf{r}})}-\Psi^\dagger(\overline{{\mathbf{r}}}) \Psi(\overline{{\mathbf{r}}})$. The linear term vanishes when we take the average and we get
\begin{equation}
\rho_Z=\overline{\rho({\mathbf{r}})}-\Psi^\dagger(\overline{\mathbf{r}}) \Psi(\overline{{\mathbf{r}}})= \frac{1}{2}\frac{\partial^2[\Psi^\dagger(\overline{\mathbf{r}}) \Psi(\overline{\mathbf{r}})]}{\partial r_i \partial r_j} ~\delta r_i\delta r_j
\end{equation}

The amplitude of the oscillations can be estimated to be of the order of the Compton wavelength, i.e.,
\begin{equation}
\overline {\delta{\mathbf{r}}^2} \sim \frac{\hbar^2}{m^2c^2}.
\end{equation}
Using the above estimate and the fact that the Zitterbewegung is isotropic, we obtain
\begin{equation}
\rho_Z\sim\frac{\hbar^2}{6m^2c^2}~\delta_{ij}~\frac{\partial^2[\Psi^\dagger (\overline{\mathbf{r}})\Psi(\overline{{\mathbf{r}}})]}{\partial r_i \partial r_j}=\frac{\hbar^2}{6m^2c^2}~\Delta(\Psi^\dagger\Psi).
\end{equation}

This yields a polarization density:
\begin{equation}
\mathbf{P}_Z = -\frac{q\hbar^2}{6m^2c^2}~\nabla(\Psi^\dagger\Psi),
\end{equation}
which is to be compared with Eq. (\ref{Eq:darwin_pol}). The functional dependence is correct, and even the multiplicative constant only differs by a factor equal to $3/4$. This is of course due to the crude estimate for the amplitude of the fluctuations.

{\bf
To conclude this section, we recall that a similar partition of the density and current into a free and a bound part can be formally obtained through a Gordon decomposition \cite{Strange}. The latter is an exact result that re-expresses the 4-current as the sum of an external ("free") and an internal ("bound") contributions. It can further be shown that the internal part may be written in terms of a polarization and magnetization density. Nevertheless, the Gordon decomposition is a formal procedure that relies on the Dirac bi-spinors. In order to recover our result (based on two-component Pauli spinors), one should apply a Foldy-Wouthuysen transformation on the bi-spinor, which is not an easy task and has been done only for the lower order terms.
Further, our approach automatically couples the equation of motion (Dirac) to the field equations (Maxwell), paving the way to the treatment of the coupled self-consistent Dirac-Maxwell system (see Sec. \ref{Dirac-Maxwell}).
}

\section{Continuity Equation}
Any meaningful description of a system of charged particles must obey a conservation law, which is usually written in the form of a continuity equation. Indeed, Maxwell's equations implicitly contain such a continuity equation and therefore automatically satisfy the conservation of electric charge.

Here, the charge and current densities that we have found at second order in $1/c$ constitute the sources of a self-consistent Dirac-Maxwell theory at the same order which we are trying to develop. Therefore, they must satisfy the continuity equation
\begin{equation}
\partial_t\rho+\nabla\cdot \mathbf{j}=0.
\end{equation}
It is easy to show that the same equation must be satisfied by the {\em free} density and current, i.e.
\begin{equation}
\partial_t\rho^{free}+\nabla\cdot \mathbf{j}^{free}=0.
\end{equation}
Using the definitions of the free sources, Eqs. (\ref{Eq:rho_free})-(\ref{Eq:j_free}), we obtain the following continuity equation:
\begin{eqnarray}
0=\partial_t(\Psi^\dagger\Psi)+\frac{i\hbar}{2m}[(\Delta\Psi^\dagger)\Psi- \Psi^\dagger(\Delta\Psi)]-\frac{q}{m}\nabla\cdot(\mathbf{A}\Psi^\dagger\Psi)+ \frac{q\hbar}{4m^2c^2}\nabla\cdot[\mathbf{E}\times (\Psi^\dagger\boldsymbol{\sigma}\Psi)], \label{Eq:Char_Cons}
\end{eqnarray}
where the only second order correction comes from the free current.

The above continuity equation refers to the conservation of charge dictated by the Maxwell equations. But this conservation law should be compatible with the relevant equation of motion, i.e. the EPE. In order to check this compatibility, we will next derive the continuity equation either from the Lagrangian density or directly from the EPE, and show that both methods yield the same result as Eq. (\ref{Eq:Char_Cons}).

\subsection{Continuity Equation from the Extended Pauli Equation}
This method involves manipulating the Hamiltonian of the system.
From the evolution equation
\begin{equation}
i\hbar\partial_t\Psi=\hat{H}\Psi,
\end{equation}
one can easily deduce that:
\begin{equation}
\partial_t(\Psi^\dagger\Psi)
=\frac{1}{i\hbar}[\Psi^\dagger(\hat{H}\Psi)-(\hat{H}^\dagger\Psi^\dagger)\Psi], \label{Eq:Con_Paul}
\end{equation}
where $\hat{H}$ is the extended Pauli Hamiltonian [Eq. (\ref{Eq:hamiltonian})] where we have neglected the kinetic energy correction ($\mathbf{p}^4$ term), and $\hat{H}^\dagger$ is its Hermitian conjugate.

We obtain:
\begin{equation}
\begin{aligned}
i\hbar\partial_t(\Psi^\dagger\Psi)&=\Psi^\dagger(\hat{H}\Psi)- (\hat{H}^\dagger\Psi^\dagger)\Psi\\
&=\Psi^\dagger\left[\frac{1}{2m}\hat{\mathbf{p}}^2\Psi-\frac{1}{m}q\mathbf{A} \cdot\hat{\mathbf{p}}\Psi-\frac{q\hbar}{4m^2c^2}\boldsymbol{\sigma} \cdot\{(\nabla\phi+ \partial_t\mathbf{A})\times\hat{\mathbf{p}}\Psi\}\right]\\
&-\left[\frac{1}{2m}\hat{\mathbf{p}}^2\Psi^\dagger+\frac{1}{m}q\mathbf{A} \cdot\hat{\mathbf{p}}\Psi^\dagger+\frac{q\hbar}{4m^2c^2}\{(\nabla\phi+\partial_t \mathbf{A})\times\hat{\mathbf{p}}\Psi^\dagger\}\cdot\boldsymbol{\sigma}\right]\Psi\\
&+\Psi^\dagger\left[-\frac{1}{m}\hat{\mathbf{p}}\cdot(q\mathbf{A})- \frac{q\hbar}{4m^2c^2}\boldsymbol{\sigma}\cdot(\hat{\mathbf{p}} \times\partial_t\mathbf{A})\right]\Psi.
\end{aligned}
\end{equation}

The last term $\hat{\mathbf{p}}\times\partial_t\mathbf{A}$ can be written as $\hat{\mathbf{p}}\times(\nabla\phi+\partial_t\mathbf{A})$ since the curl of a gradient is zero. Rearranging the terms, we obtain the same continuity equation as in Eq. (\ref{Eq:Char_Cons}).

However, we stress that by using this method the actual sources cannot be obtained, because the magnetization and polarization terms cancel each other out in the continuity equation. The Lagrangian procedure detailed in the preceding sections was therefore necessary to derive the actual sources that should go into the Maxwell equations.

\subsection{Continuity Equation from Noether's Theorem}
Noether's theorem holds a special place in Lagrangian mechanics. It states that we have a conserved quantity whenever there is a symmetry in the system (invariance under some type of transformation). In the present case, the relevant symmetry is gauge invariance and the corresponding conserved quantity is the electric charge.

Noether's theorem allows us to calculate the continuity equation directly from the Lagrangian density and can be written in a four-vector form as follwos
\begin{equation}
\partial_\mu j^\mu= \partial_\mu\left(\Psi^\dagger\frac{\partial\mathcal{L}} {\partial(\partial_\mu\Psi^\dagger)}- \frac{\partial\mathcal{L}}{\partial(\partial_\mu\Psi)}\Psi\right)=0
\label{Eq:noether}
\end{equation}
where $\partial_\mu=\{\partial_{ct},\nabla\}$ and the Einstein summation convention is used. For the continuity equation, only the terms depending on the derivatives of $\Psi$ and $\Psi^\dagger$ are relevant. Thus, we rewrite the Lagrangian density as
\begin{equation}
\begin{aligned}
\mathcal{L}=&\mathcal{L}^\prime+\frac{i\hbar}{2}(\Psi^\dagger\dot{\Psi}-\dot{\Psi^\dagger}\Psi)-\frac{1}{2m}[i\hbar\partial_k\Psi^\dagger(-i\hbar\partial_k-q A_k)\Psi+i\hbar qA_k\Psi^\dagger \partial_k\Psi]\\
&-\frac{q\hbar}{8m^2c^2}\epsilon_{ijk}[\Psi^\dagger\sigma_i\partial_j\phi\hat{p}_k\Psi-\partial_j\phi\hat{p}_k\Psi^\dagger\sigma_i \Psi +\Psi^\dagger\sigma_i\partial_t A_j\hat{p}_k\Psi-\partial_tA_j \hat{p}_k \Psi^\dagger\sigma_i\Psi],
\end{aligned}
\end{equation}
where $\mathcal{L}^\prime$ contains all the remaining terms of the Lagrangian density.

The time components yields
\begin{equation}
\partial_{ct}\left[\Psi^\dagger\frac{\partial\mathcal{L}} {\partial(\partial_{ct}\Psi^\dagger)}-\frac{\partial\mathcal{L}} {\partial(\partial_{ct}\Psi)}\Psi\right]=-i\hbar\partial_t(\Psi^\dagger\Psi).
\label{Eq:noether-time}
\end{equation}
For the space components, we have
\begin{eqnarray}
\Psi^\dagger\frac{\partial\mathcal{L}}{\partial(\partial_k\Psi^\dagger)}&=& \frac{i\hbar}{2m}\Psi^\dagger(i\hbar\partial_k+qA_k)\Psi-\frac{q\hbar(i\hbar)}{8m^2c^2} \epsilon_{ijk}(\partial_j\phi+\partial_tA_j)\Psi^\dagger\sigma_i\Psi\\
-\frac{\partial\mathcal{L}}{\partial(\partial_k\Psi)}\Psi&=& -\frac{i\hbar}{2m}(i\hbar\partial_k\Psi^\dagger-qA_k)\Psi- \frac{q\hbar(i\hbar)}{8m^2c^2}\epsilon_{ijk}(\partial_j\phi+ \partial_tA_j)\Psi^\dagger\sigma_i\Psi.
\end{eqnarray}
Combining the above terms yields
\begin{equation}
\begin{aligned}
\partial_k\left[\Psi^\dagger\frac{\partial\mathcal{L}}{\partial(\partial_k\Psi^\dagger)}- \frac{\partial\mathcal{L}}{\partial(\partial_k\Psi)}\Psi\right]= &~\partial_k\frac{(i\hbar)^2}{2m}\{\Psi^\dagger(\partial_k\Psi)- (\partial_k\Psi^\dagger)\Psi\}+\frac{qi\hbar}{m}\partial_k(A_k\Psi^\dagger\Psi)\\
&-\frac{q\hbar(i\hbar)}{4m^2c^2}\epsilon_{ijk}\partial_k[(\partial_j\phi+\partial_tA_j) \Psi^\dagger\sigma_i\Psi]\\
=&-\frac{(i\hbar)^2}{2m}[(\Delta\Psi^\dagger)\Psi-\Psi^\dagger(\Delta\Psi)]+ \frac{qi\hbar}{m}\nabla\cdot(\mathbf{A}\Psi^\dagger\Psi)\\
&+\frac{q\hbar(i\hbar)}{4m^2c^2}\nabla\cdot[(\nabla\phi+ \partial_t\mathbf{A})\times(\Psi^\dagger\boldsymbol{\sigma}\Psi)]\\
\end{aligned}\label{eq:noether_curr}
\end{equation}

Finally, substituting Eqs. (\ref{eq:noether_curr}) and (\ref{Eq:noether-time}) into Eq. (\ref{Eq:noether}), we arrive again at the same continuity equation (\ref{Eq:Char_Cons}).

Again, we stress that the full density and current corrections could not have been obtained from Noether's theorem, as the latter only provides the conservation law (continuity equation), but not the sources in themselves.
That the conservation law obtained from the Pauli Hamiltonian or Noether's theorem actually coincides with the charge conservation that is implicit in Maxwell's equations with the second-order sources is an attractive feature of our model and strengthens our confidence in its validity.

\section{Coupled Dirac-Maxwell equations}\label{Dirac-Maxwell}
So far, we developed and described a model for obtaining the sources and the equation of motion by dint of a Lagrangian approach.
The resulting equation of motion (extended Pauli equation, EPE) is a second order (in $1/c$) approximation for the positive energy states of the Dirac equation.

However, there is some inconsistency here. The charge and current density were derived by putting together the Pauli and Maxwell parts of the Lagrangian, i.e., by assuming that the EPE acts as some sort of source to the Maxwell equations. But the latter are exact to all orders in $1/c$, whereas the EPE is only valid at second order. While this procedure yields the correct result as far as the currents and densities are concerned, it would be desirable to construct a model that treats on the same footing (i.e., at the same order) both the equation of motion (Pauli) and the equations for the fields (Maxwell).

It has been known for a long time \cite{LBLL,Holland-Brown,Rousseaux,Montigny2006} that the Maxwell equations possess two independent non-relativistic limits, which correspond to situations where either $|{\mathbf E}| \gg c|{\mathbf B}|$ (electric limit) or $|{\mathbf E}| \ll c|{\mathbf B}|$ (magnetic limit). Each of the two limits is Galilei covariant. In practice, the electric limit amounts to neglecting the time-derivative of the magnetic field in Faraday's law of induction, whereas the magnetic limit is obtained by dropping the displacement current in Amp\`{e}re's equation.

Both limits can be derived in a rigorous and gauge independent way by using a non-dimensional version of Maxwell equations \cite{manfredi_maxwell}. Here, we present a short derivation that uses SI units and the Lorentz gauge.
In the present case, we are concerned with the electric limit (the magnetic limit is by construction charge neutral and thus requires the presence of two mobile species of opposite charge).

The electric limit can be viewed as the case where $c \to \infty$ while $\varepsilon_0$ remains finite. We rewrite the Maxwell equations as:
\begin{eqnarray}
-\Delta \phi + \frac{1}{c^2}\frac{\partial^2 \phi}{\partial t^2} &=& \frac{q\rho}{\varepsilon_0}, \\
-\Delta \mathbf{A} + \frac{1}{c^2}\frac{\partial^2 \mathbf{A}}{\partial t^2} &=& \frac{q\mathbf{j}}{\varepsilon_0 c^2}~.
\end{eqnarray}
We then expand all variables in powers of $c^{-2}$: $\phi=\phi_0+\phi_2+\dots$, $\mathbf{A}=\mathbf{A}_0+\mathbf{A}_2+\dots$, and likewise for $\rho$ and $\mathbf{j}$. We obtain, at zeroth and second order:
\begin{eqnarray}
\Delta \mathbf{A}_0 &=& 0 \label{Eq:max_A0}\\
-\Delta \phi_0  &=& \frac{q\rho_0}{\varepsilon_0}, \label{Eq:max_phi0}\\
-\Delta \phi_2 &+& \frac{1}{c^2}\frac{\partial^2 \phi_0}{\partial t^2} = \frac{q\rho_2}{\varepsilon_0}~, \label{Eq:max_phi2}\\
-\Delta \mathbf{A}_2 &=& \frac{q\mathbf{j}_0}{\varepsilon_0 c^2}~. \label{Eq:max_A2}
\end{eqnarray}
If we assume zero boundary conditions at infinity, we have that $\mathbf{A}_0 =0$: there is no magnetic field at leading order, as is natural in the electric limit.

In terms of the electric and magnetic fields, we have at zeroth order:
\begin{eqnarray}
\nabla \cdot \mathbf{E}_0 &=& \frac{q\rho_0}{\varepsilon_0}, \\
\nabla \times \mathbf{E}_0 &=&\nabla \cdot \mathbf{B}_0  =\nabla \times \mathbf{B}_0,
\end{eqnarray}
so that $\mathbf{B}_0 =0$ and $\mathbf{E}_0 =-\nabla \phi_0$, which immediately yields Eqs. (\ref{Eq:max_A0})-(\ref{Eq:max_phi0}). At second order, one finds
\begin{eqnarray}
\nabla \cdot \mathbf{E}_2 &=& \frac{q\rho_2}{\varepsilon_0}, \\
\nabla \times \mathbf{E}_2 &=& \frac{\partial \mathbf{B}_2}{\partial t} \\
\nabla \cdot \mathbf{B}_2 &=& 0 \\
\nabla \times \mathbf{B}_2 &=& \frac{q\mathbf{j}_0}{\varepsilon_0 c^2} +
 \frac{1}{c^2}\frac{\partial \mathbf{E}_0}{\partial t}. \label{Eq_Max_B2}
\end{eqnarray}
Using $\mathbf{E}_2 = -\nabla\phi_2-\partial_t \mathbf{A}_2$ and the Lorentz gauge condition
\begin{equation}
\nabla \cdot \mathbf{A}_2+ \frac{1}{c^2}\frac{\partial\phi_0}{\partial t}=0,
\end{equation}
we obtain Eqs. (\ref{Eq:max_phi2})-(\ref{Eq:max_A2}).

By taking the divergence of Eq. (\ref{Eq_Max_B2}), we obtain the continuity equation at zeroth order:
\begin{equation}
\frac{\partial \rho_0}{\partial t} + \nabla \cdot \mathbf{j}_0= 0. \end{equation}
The continuity equation at second order
\begin{equation}
\frac{\partial \rho_2}{\partial t} + \nabla \cdot \mathbf{j}_2= 0. \end{equation}
can likewise be obtained by pursuing the expansion to fourth order.

We can now try to match the above sources with those that we found with the general Lagrangian approach, i.e. Eqs. (\ref{Eq:rho_free})-(\ref{Eq:darwin_pol}). For the charge density, it is clear that $\rho_0$ coincides with the free density, while $\rho_2$ can be identified with the bound density. For the current things are subtler, as the free current contains zeroth-order as well as second-order terms. We have
\begin{eqnarray}
q\mathbf{j}_0&=& \underbrace{\frac{i\hbar}{2m}[(\nabla\Psi^\dagger)\Psi- \Psi^\dagger(\nabla\Psi)]}_{q\mathbf{j}^{free}_0}
 + \underbrace{\nabla\times \mathbf{M}}_{q\mathbf{j}^{bound}_0} \label{eq:j0}\\
q\mathbf{j}_2 &=& \underbrace{-\frac{q}{m}\mathbf{A}_2\Psi^\dagger\Psi -
\frac{q\hbar}{4m^2c^2} [\nabla\phi_0\times(\Psi^\dagger\boldsymbol{\sigma}\Psi)]}_{q\mathbf{j}^{free}_2} +
\underbrace{\partial_t \mathbf{P}}_{q\mathbf{j}^{bound}_2}, \label{eq:j2}
\end{eqnarray}
where we have indicated explicitly the order of the potentials. Further, note that the term depending on the vector potential in $\mathbf{P}_{spin}$ [Eq. (\ref{Eq:spin_pol})] can be neglected because it is of higher order.

In principle, only the current $\mathbf{j}_0$ should go into Amp\`{e}re's equation (\ref{Eq:max_A2}). However, by doing so the continuity equation (\ref{Eq:Char_Cons}) would no longer be satisfied, which is an undesirable property. For this reason, we shall keep both the zeroth and second order currents in Amp\`{e}re's equation, even though this introduces some spurious fourth-order terms. By doing so, the Maxwell equations become:
\begin{eqnarray}
\mathbf{A}_0 &=& 0\\
-\Delta \phi_0  &=& \frac{q\rho_0}{\varepsilon_0}, \label{Eq:max_phi_fin}\\
-\Delta \phi_2 &+& \frac{1}{c^2}\frac{\partial^2 \phi_0}{\partial t^2} = \frac{q\rho_2}{\varepsilon_0}~, \\
-\Delta \mathbf{A}_2 &=& \frac{q}{\varepsilon_0 c^2}~(\mathbf{j}_0+\mathbf{j}_2). \label{Eq:max_A_fin}
\end{eqnarray}

We are now able to write down a family of useful models that can be viewed as self-consistent expansions of the original Dirac-Maxwell equations to second order in $1/c$.

\subsection{Purely internal electromagnetic fields}
If we assume that the electromagnetic fields are purely internal (i.e., self-consistent) and take into account that $\mathbf{A}_0=0$, we obtain the following extended Pauli Hamiltonian (we neglect the constant $mc^2$):
\begin{equation}\label{Eq:Red_Pauli}
\hat{H}=\,q(\phi_0+\phi_2)+ \frac{\hat{\mathbf{p}}^2}{2m}- \frac{q}{2m}(\hat{\mathbf{p}}\cdot\mathbf{A}_2+ \mathbf{A}_2\cdot\hat{\mathbf{p}}) + \frac{q\hbar}{2m}\boldsymbol{\sigma}\cdot \nabla\times\mathbf{A}_2
-\frac{q\hbar^2}{8m^2c^2}\Delta\phi_0- \frac{q\hbar}{4m^2c^2}\boldsymbol{\sigma}\cdot\nabla\phi_0\times\hat{\mathbf{p}},
\end{equation}
which is to be coupled to the reduced Maxwell equations (\ref{Eq:max_phi_fin})-(\ref{Eq:max_A_fin}).

Using Noether's theorem or manipulating the corresponding EPE yields the continuity equation as:
\begin{equation}\label{Eq:Red_Cont_Noe}
0=\partial_t(\Psi^\dagger\Psi)+ \frac{i\hbar}{2m}[(\Delta\Psi^\dagger)\Psi-\Psi^\dagger(\Delta\Psi)]- \frac{q}{m}\nabla\cdot(\mathbf{A}_2\Psi^\dagger\Psi)+ \frac{q\hbar}{4m^2c^2}\nabla\cdot[\nabla\phi_0 \times(\Psi^\dagger\boldsymbol{\sigma}\Psi)],
\end{equation}
which is consistent with
\begin{equation}
\frac{\partial (\rho_0 +\rho_2)}{\partial t} + \nabla \cdot (\mathbf{j}_0+\mathbf{j}_2)= 0,
\end{equation}
the currents being defined as in Eq. (\ref{eq:j0})-(\ref{eq:j2}).

\subsection{Internal and external electromagnetic fields}
If some external electromagnetic fields are also present (e.g., laser pulse), these can be assumed to be of zeroth order. For the scalar potential, nothing needs to be changed, as $\phi$ already contains a zeroth order term: one just rewrites $\phi_0 = \phi_0^{int} +  \phi_0^{ext}$.

In contrast, for the vector potential we now have a zeroth order term $\mathbf{A}_0=\mathbf{A}_0^{ext}$ and cannot anymore make the simplifications leading to Eq. (\ref{Eq:Red_Pauli}). Thus we stick with the full Hamiltonian of Eq. (\ref{Eq:hamiltonian_pot}) (although in the second-order terms one can neglect the self-consistent vector potential $\mathbf{A}_2$)
and the corresponding continuity equation (\ref{Eq:Char_Cons}). The sources to be used in the reduced Maxwell equations are those of Eqs. (\ref{Eq:rho_free})-(\ref{Eq:darwin_pol}).

\subsection{Minimal self-consistent model}
We have seen that self-consistent electric effects appear at zeroth order in the Maxwell equations, whereas magnetic effects are of second order. It may then be reasonable (if not mathematically rigorous) to neglect self-consistent electrostatic corrections at second order. We shall see that this entails significant simplifications in the resulting equations.

The relevant Hamiltonian becomes, neglecting $\phi_2$ in Eq. (\ref{Eq:Red_Pauli}):
\begin{equation}\label{Eq:Red_Pauli2}
\hat{H}=\,q\phi_0+ \frac{\hat{\mathbf{p}}^2}{2m}- \frac{q}{2m}(\hat{\mathbf{p}}\cdot\mathbf{A}_2+ \mathbf{A}_2\cdot\hat{\mathbf{p}}) + \frac{q\hbar}{2m}\boldsymbol{\sigma}\cdot \nabla\times\mathbf{A}_2
-\frac{q\hbar^2}{8m^2c^2}\Delta\phi_0- \frac{q\hbar}{4m^2c^2}\boldsymbol{\sigma}\cdot\nabla\phi_0\times\hat{\mathbf{p}},
\end{equation}
whereas for the Maxwell equations we have
\begin{eqnarray}
-\Delta \phi_0 &=& \frac{q\rho_0}{\varepsilon_0}, \label{Eq:max_phi3}\\
-\Delta \mathbf{A}_2 &=& \frac{q}{\varepsilon_0 c^2}~(\mathbf{j}_0+\mathbf{j}_2^{free}). \label{Eq:max_A3}
\end{eqnarray}
We see that the Maxwell equations now reduce to a set of two elliptic (i.e., Poisson-like) equations where no time derivatives appear.
The last term in the current is needed to ensure that the continuity equation derived from the Hamiltonian (\ref{Eq:Red_Pauli2}) [i.e. Eq. (\ref{Eq:Red_Cont_Noe})] is consistent with
\begin{equation}
\frac{\partial \rho_0}{\partial t} + \nabla \cdot (\mathbf{j}_0+\mathbf{j}_2^{free})= 0.
\end{equation}

Another welcome property of the above set of equations is that the particle density is simply defined as $\rho_0=\Psi^\dagger\Psi$, as in the Schr\"{o}dinger or Pauli equations.

Finally, one can show that the self-consistent vector potential $\mathbf{A}_2$, when re-injected into the Hamiltonian (\ref{Eq:Red_Pauli2}), yields all the terms that are present in the Breit equation for a many-electron system in the mean-field approximation (spin-spin, spin-same-orbit, spin-other-orbit interactions, etc \dots). Thus, a lot of physically important information is already captured by such a simplified model. This issue will be investigated in a future work.

\section{Conclusion}
Relativistic effects can have an impact on the electron dynamics in heavy atoms, dense plasmas, and condensed-matter systems excited with intense and ultrafast laser pulses. In particular, the electron spin can couple not only to the electric field of the static nuclei (this is the ordinary spin-orbit coupling), but also to the self-consistent mean field generated by all other electrons, or directly to the magnetic and electric fields of the incident laser pulse. In view of this complex variety of possible physical mechanisms, it may be necessary to go beyond the lowest order description of an electron with spin, i.e. the Pauli equation, sometimes supplemented by an ad-hoc spin-orbit term.

The purpose of this paper was to derive a self-consistent mean-field model that incorporates quantum, spin, and relativistic effects up to second order in $1/c$. We started from the second-order Hamiltonian derived from the Dirac equation through a Foldy-Wouthuysen transformation. We called the corresponding equation of motion the "extended Pauli equation" (EPE). Then we constructed a Lagrangian density that reproduces the EPE via the Euler-Lagrange equations.

In order to couple the EPE to the Maxwell equations, we added the standard electromagnetic Lagrangian to the Lagrangian for the EPE. The advantage of this approach is that, when recovering the Maxwell equations from the full Lagrangian density through the Euler-Lagrange equations, one automatically gets the expressions for the sources (charge and current densities) up to second order.

A physical interpretation was given for each new term appearing in the sources.
At zeroth order, one recovers the standard expressions for the Schr\"{o}dinger density and current. The second-order sources contain a free current correction, as well as other terms that can be written in the form of a polarization and a magnetization density. The magnetization is linked to the divergence-free ``spin current" that already appears in the standard Pauli equation. The polarization terms can be split into a ``Darwin" part and a ``spin" part. The former originates from the relativistic Zittebewegung (fast oscillations of the electron trajectory around its mean value), which causes the density to smear out on a distance of the order of the Compton wavelength.
The spin polarization was interpreted as a Lorentz transformation of the magnetization density in the rest frame of the electron.

The charge density and current were derived by putting together the Pauli and the Maxwell parts of the Lagrangian, i.e., by assuming that the EPE acts as some sort of source to the Maxwell equations. But the latter are exact to all orders in $1/c$, whereas the EPE is only valid up to second order. In order to
treat on the same footing (i.e., at the same order) both the equation of motion (Pauli) and the equations for the fields (Maxwell), we also expanded the Maxwell equations to second order in $1/c$. With this procedure, we were able to construct a fully self-consistent set of equations that are valid at second order and respect some appropriate conservation laws. Several versions of such a model were discussed.

The models that we have derived should be useful, for instance, for applications to dense and weakly degenerate electron plasmas created via intense laser pulses. Energetic electrons are routinely observed in such plasmas and their relativistic velocities can be used to modify the properties of incident radiation \cite{Kiefer}. Other possible areas of applications involve inertial confinement fusion \cite{Haines} and astrophysical plasmas \cite{Ridgers}, as well as nanometric systems (nanoparticles, thin films) excited with ultrashort laser pulses in the femto- or attosecond domain \cite{Bigot_natphys}.

Finally, it should be noted that the models discussed in this paper are limited to the mean-field approximation, although, in contrast to most other such models, the {\em magnetic} mean field is also included. Going beyond the mean-field approximation would require appropriate exchange and correlation functionals \cite{Engel, Strange}, which are notoriously difficult to obtain in the relativistic regime. Their semi-relativistic expansion and compatibility with the reduced Maxwell equations will also need to be examined.

\section{Acknowledgements}
This work was partially funded by the European Research Council Advanced grant ``Atomag" (ERC-2009-AdG-20090325 247452). We thank Dr. Jean-Yves Bigot for his scientific and financial support.
One of the authors (A. Dixit) would like to thank Aleksandra Pasieczna for her personal support.

\pagebreak

\appendix
\section{Relativistic mass correction} \label{App:Rel_Corr}
To include the relativistic mass correction term to the Lagrangian density, we use the form $\bra{\Psi}\hat{\mathbf{p}}^2\hat{\mathbf{p}}^2\ket{\Psi}$. This is similar to the way the term $(\hat{\mathbf{p}}-q\mathbf{A})^2/2m$ is used in the original Lagrangian density, i.e., $\bra{\Psi}(\hat{\mathbf{p}}-q\mathbf{A})\cdot(\hat{\mathbf{p}}-q\mathbf{A})\ket{\Psi}$. We use the following highly reduced Pauli equation to compute the contribution of this term:
\begin{equation}
i\hbar\partial_t\Psi=\frac{(\hat{\mathbf{p}}-q\mathbf{A})^4}{8m^3c^2}\Psi.
\end{equation}
The Lagrangian density for this Pauli equation coupled to the Maxwell equations is
\begin{equation}
\begin{aligned}
\mathcal{L}=&\frac{i\hbar}{2}(\Psi^\dagger\dot{\Psi}-\dot{\Psi^\dagger}\Psi)+ \frac{1}{8m^3c^2}(-\hat{p}_i-qA_i) (-\hat{p}_i-qA_i)\Psi^\dagger(\hat{p}_j-qA_j)(\hat{p}_j-qA_j)\Psi\\
&+\frac{\varepsilon_0}{2}(\partial_k\phi)^2-\frac{\varepsilon_0}{2c^2} (\partial_t\phi)^2-\frac{1}{2\mu_0}(\partial_jA_k)^2+\frac{1}{2\mu_0 c^2}(\partial_t A_k)^2.
\end{aligned}
\end{equation}

One can easily check that the probability density will not change. Thus, we rewrite the Lagrangian density as one consisting of terms containing only the Maxwell terms for the vector field and the relativistic correction:
\begin{equation}
\mathcal{L}=\mathcal{L}^\prime-\frac{1}{2\mu_0}(\partial_jA_k)^2+\frac{1}{2\mu_0 c^2}(\partial_t A_k)^2+\frac{1}{8m^3c^2}(\hat{p}_i+qA_i)(\hat{p}_i+qA_i)\Psi^\dagger(\hat{p}_j-qA_j) (\hat{p}_j-qA_j)\Psi,
\end{equation}
where $\mathcal{L}^\prime$ contains all remaining terms. The momentum operator can act on both the vector potential and the wavefunction. Simplifying and putting the $\hat{p}_i\hat{p}_i\Psi^\dagger\hat{p}_j\hat{p}_j\Psi$ term in $\mathcal{L}^\prime$, gives
\begin{equation}
\begin{aligned}
\mathcal{L}&=\mathcal{L}^\prime-\frac{1}{2\mu_0}(\partial_jA_k)^2+\frac{1}{2\mu_0 c^2}(\partial_t A_k)^2+\frac{1}{8m^3c^2}[\hat{p}_i\hat{p}_i\Psi^\dagger q^2A_jA_j\Psi+q^2A_iA_i\Psi^\dagger\hat{p}_j\hat{p}_j\Psi]\\
&+\frac{1}{8m^3c^2}\left[\{2qA_i\hat{p}_i+(\hat{p}_iqA_i)+ q^2A_iA_i\}\Psi^\dagger\{q^2A_jA_j-2qA_j\hat{p}_j-(\hat{p}_jqA_j)\}\Psi\right].
\end{aligned}
\end{equation}

The Euler-Lagrange equations for $A_k$ will give the corresponding component of the current density. However, we must be careful when we calculate these terms, since there are two indices $i$ and $j$. Computing for a single component (say, the $k^{th}$ component) will give a $\delta_{ik}$ and $\delta_{jk}$. We obtain
\begin{equation}
\begin{aligned}
\frac{\partial\mathcal{L}}{\partial A_k}=&\frac{1}{8m^3c^2}\left[8q^2(q\mathbf{A}\cdot q\mathbf{A})\Psi^\dagger\Psi A_k+4i\hbar q^2\left\{q\mathbf{A}\cdot[\Psi^\dagger(\nabla\Psi)- (\nabla\Psi^\dagger)\Psi]\right\}A_k\right]\\
&+\frac{1}{8m^3c^2}\left[-2qi\hbar (q\mathbf{A}\cdot q\mathbf{A})[(\partial_k\Psi^\dagger)\Psi-\Psi^\dagger(\partial_k\Psi)]- 2(qi\hbar)^2(\nabla\cdot\mathbf{A})\partial_k(\Psi^\dagger\Psi)\right]\\
&+\frac{1}{8m^3c^2}\left[-4(qi\hbar)^2\left\{(\partial_k\Psi^\dagger) (\mathbf{A}\cdot\nabla\Psi)+(\mathbf{A}\cdot\nabla\Psi^\dagger) (\partial_k\Psi)\right\}\right]\\
&+\frac{1}{8m^3c^2}2(qi\hbar)^2\{A_k\Psi^\dagger(\Delta\Psi)+ (\Delta\Psi^\dagger)A_k\Psi\}.
\end{aligned}
\end{equation}

For the time derivative components of the Euler-Lagrange equations, we have
\begin{equation}
\partial_t\frac{\partial\mathcal{L}}{\partial (\partial_tA_k)}=\frac{1}{\mu_0 c^2}\partial_t^2A_k.
\end{equation}
For the spatial derivatives, when we calculate the components with the index $i$, we replace $j$ by $i$. Thus,
\begin{equation}
\begin{aligned}
\frac{\partial\mathcal{L}}{\partial (\partial_jA_k)}=&-\frac{1}{\mu_0}\partial_jA_k+\frac{1}{8m^3c^2}\delta_{jk}\left[-qi\hbar\Psi^\dagger\{q^2(\mathbf{A}\cdot\mathbf{A})+2qi\hbar\mathbf{A}\cdot\nabla+qi\hbar(\nabla\cdot\mathbf{A})\}\Psi\right]\\
&+\frac{1}{8m^3c^2}\delta_{jk}\left[\{q^2\mathbf{A}\cdot\mathbf{A}-2qi\hbar\mathbf{A}\cdot\nabla-qi\hbar(\nabla\cdot\mathbf{A})\}\Psi^\dagger qi\hbar\Psi\right].
\end{aligned}
\end{equation}
Simplifying, one obtains
\begin{equation}
\begin{aligned}
\partial_j\frac{\partial\mathcal{L}}{\partial (\partial_jA_k)}=&-\frac{(qi\hbar)^2}{4m^3c^2}\partial_k\left[\{\Psi^\dagger(\mathbf{A}\cdot\nabla\Psi)+ (\mathbf{A}\cdot\nabla\Psi^\dagger)\Psi\}+(\nabla\cdot\mathbf{A})\Psi^\dagger\Psi\right]\\
&-\frac{1}{\mu_0}\Delta A_k
\end{aligned}
\end{equation}
Substituting these terms in the Euler-Lagrange equation and rearranging:
\begin{equation}
\begin{aligned}
{8m^3c^2}\left\{-\frac{1}{\mu_0}\Delta A_k+\frac{1}{\mu_0 c^2}\partial_t^2A_k\right\}=& ~ 2(qi\hbar)^2\partial_k\nabla\cdot(\mathbf{A}\Psi^\dagger\Psi)+8q^2(q\mathbf{A}\cdot q\mathbf{A})\Psi^\dagger\Psi A_k\\
&+4i\hbar q^2\left\{q\mathbf{A}\cdot[\Psi^\dagger(\nabla\Psi)-(\nabla\Psi^\dagger)\Psi]\right\}A_k\\
&-2qi\hbar (q\mathbf{A}\cdot q\mathbf{A})[(\partial_k\Psi^\dagger)\Psi-\Psi^\dagger(\partial_k\Psi)]\\
&-2(qi\hbar)^2[(\partial_k\Psi^\dagger)\{\nabla\cdot(\mathbf{A}\Psi)\}+\{\nabla\cdot(\mathbf{A}\Psi^\dagger)\}(\partial_k\Psi)]\\
&-2(qi\hbar)^2[(\mathbf{A}\cdot\nabla\Psi^\dagger)(\partial_k\Psi)+(\partial_k\Psi^\dagger)(\mathbf{A}\cdot\nabla\Psi)]\\
&+2(qi\hbar)^2\{A_k\Psi^\dagger(\Delta\Psi)+(\Delta\Psi^\dagger)A_k\Psi\}\\
&= 8m^3c^2qj_k.
\end{aligned}
\end{equation}
Converting to vector form and rewriting the current, we get the relativistic mass correction to the current density
\begin{equation}
\begin{aligned}
\mathbf{j}^{rel}&=\frac{q\hbar^2}{4m^3c^2}[(\nabla\Psi^\dagger)\nabla\cdot(\mathbf{A} \Psi)+\nabla\cdot(\mathbf{A}\Psi^\dagger)(\nabla\Psi)+ (\mathbf{A}\cdot\nabla\Psi^\dagger)(\nabla\Psi)+(\nabla\Psi^\dagger)(\mathbf{A}\cdot\nabla\Psi)]\\
&\quad{}-\frac{q\hbar^2}{4m^3c^2}\nabla\{\nabla\cdot(\mathbf{A}\Psi^\dagger\Psi)\}+ \frac{q^3}{m^3c^2}(\mathbf{A}\cdot\mathbf{A})(\Psi^\dagger\mathbf{A}\Psi)+\frac{i\hbar q^2\mathbf{A}}{2m^3c^2}[\Psi^\dagger(\mathbf{A}\cdot\nabla\Psi)- (\mathbf{A}\cdot\nabla\Psi^\dagger)\Psi]\\
&\quad{}-\frac{q^2i\hbar}{2m^3c^2}(\mathbf{A}\cdot \mathbf{A})[(\nabla\Psi^\dagger)\Psi- \Psi^\dagger(\nabla\Psi)]-\frac{q\hbar^2}{4m^3c^2}[\Psi^\dagger(\Delta\Psi)+ (\Delta\Psi^\dagger)\Psi]\mathbf{A}.
\end{aligned}
\end{equation}

This correction is algebraically cumbersome and its practical usefulness in numerical computations may be questioned.
It must be pointed out that all terms in $\mathbf{j}^{rel}$ depend on the vector potential at least linearly. As we have seen that the self-consistent (internal) vector potential is a second-order quantity, we conclude that, for a purely self-consistent magnetic field, $\mathbf{j}^{rel}$ does not contain any contribution at second order.
Things could be different, of course, in the case of an external magnetic field, which can be arbitrarily large.

\section{Derivation of the extended Pauli equation from the Lagrangian density} \label{App:Lagr_Test}
Here we show that the Lagrangian density provided in Eq. (\ref{Eq:Lagr_Dens}) does return the EPE when we consider the Euler-Lagrange equations for $\Psi^\dagger$.

Computing each term of the Euler-Lagrange equations yields:
\begin{equation}
\begin{aligned}
\frac{\partial\mathcal{L}}{\partial \Psi^\dagger}=&\frac{i\hbar}{2}\dot{\Psi}-(mc^2+q\phi)\Psi+ \frac{1}{2m}q\mathbf{A}\cdot(\hat{\mathbf{p}}-q\mathbf{A})\Psi \\
+& \left[\frac{q\hbar}{2m}\epsilon_{ijk}{\sigma}_i\partial_jA_k- \frac{q\hbar^2}{8m^2c^2}\partial_k^2\phi-\frac{q\hbar^2}{8m^2c^2}\partial_t\partial_k A_k\right]\Psi\\
+&\epsilon_{ijk}\left[\frac{q\hbar}{4m^2c^2}{\sigma}_i \partial_j\phi q A_k+\frac{q\hbar}{4m^2c^2}{\sigma}_i \partial_tA_j q A_k\right]\Psi-\frac{q\hbar}{8m^2c^2}\epsilon_{ijk} [\Psi^\dagger{\sigma}_i(\partial_j\phi+\partial_t A_j)\hat{p}_k\Psi],
\end{aligned}
\end{equation}
\begin{equation}
\partial_t\frac{\partial\mathcal{L}}{\partial(\partial_t\Psi^\dagger)} =-\frac{i\hbar}{2}\partial_t{\Psi},
\end{equation}
\begin{equation}
\partial_k\frac{\partial\mathcal{L}}{\partial(\partial_k\Psi^\dagger)} =\frac{1}{2m}\hat{\mathbf{p}}\cdot(\hat{\mathbf{p}}- q\mathbf{A})\Psi-\frac{q\hbar}{8m^2c^2}\hat{\mathbf{p}} \cdot[(\nabla\phi+\partial_t\mathbf{A})\times(\boldsymbol{\sigma}\Psi)].
\end{equation}

When we combine the above terms and insert them into Eq. (\ref{Eq:E-L_Eq}), we obtain, after some algebra:
\begin{equation}
\begin{aligned}
i\hbar\dot{\Psi}&=(mc^2+q\phi)\Psi+\frac{(\hat{\mathbf{p}}- q\mathbf{A})^2}{2m}\Psi-\frac{q\hbar}{2m} \boldsymbol{\sigma}\cdot(\nabla\times\mathbf{A})\Psi+ \frac{q\hbar^2}{8m^2c^2}\Delta\phi\Psi+ \frac{q\hbar^2}{8m^2c^2}\nabla\cdot\partial_t\mathbf{A}\Psi\\
&+\frac{q\hbar}{4m^2c^2}\boldsymbol{\sigma}\cdot\left\{(\nabla\phi+ \partial_t\mathbf{A})\times(\hat{\mathbf{p}}-q\mathbf{A})\right\}\Psi- \frac{q\hbar}{8m^2c^2}\boldsymbol{\sigma} \cdot(\hat{\mathbf{p}}\times\partial_t\mathbf{A})\Psi,
\end{aligned}
\end{equation}
which is the expected EPE.
The conjugate of the EPE is obtained by taking the Euler-Lagrange equation for $\Psi$.

\section{Landau's variational method}\label{App:landau}
In Ref. \cite{Landau}, Landau showed how to obtain the charge and current densities for the Pauli equation at lowest order, using a variational method. In this Appendix, we prove that Landau's method can be generalized to the EPE (second order in $1/c$) and that the resulting charge and current densities are identical to those obtained with our Lagrangian approach.

The Lagrangian method developed in the main body of this work is nevertheless much simpler to implement. The main reason is that the Lagrangian approach uses a single scalar function of the various fields, whereas Landau's method is based on the expectation value of the energy, which involves computing complicated integrals.

\subsection{Landau's result at zeroth order}
Landau considered a single electron with spin interacting with a magnetic field, described by its vector potential $\mathbf{A}$.
The starting point is to assume that an elementary variation of the expectation value of the energy can be associated to an elementary variation of the electromagnetic energy, so that one can write
\begin{equation}
\delta \left<H\right>=-\int q\mathbf{j}\cdot\delta \mathbf{A} d\tau , \label{1}
\end{equation}
where $\int d \tau=\int d^{3}\mathbf{r} \int dt$ denotes integration over space and time, and the expectation value of the energy $\left<H\right>$ is
\begin{eqnarray}
\left<H\right>&=&\int \Psi^{\dagger }H\Psi d\tau =\int \Psi^{\dagger }\left[ \frac{(\mathbf{p}-q\mathbf{A})^{2}}{2m} - \frac{q\hbar}{2m}\boldsymbol{\sigma }
\cdot \mathbf{B}\right]\Psi d\tau.   \label{2}
\end{eqnarray}

Computing the variation of the mean energy, we get:
\begin{eqnarray}
\delta \left<H\right> =\int \Psi^{\dagger }\left[ -\frac{q}{2m}\left(\mathbf{p}\cdot \delta \mathbf{A}+\delta \mathbf{A}\cdot \mathbf{p} \right)+ \frac{q^{2}}{m}\mathbf{A}\cdot \delta \mathbf{A}- \frac{qh}{2m}  \boldsymbol{\sigma }\cdot \left(\boldsymbol{\nabla} \times \delta \mathbf{A} \right)\right]\Psi d\tau .  \label{3}
\end{eqnarray}
Using $\mathbf{p}=-i\hbar \boldsymbol{\nabla}$ and integrating by parts, yields
\begin{eqnarray}
\delta \left<H\right>&=& -\int q\left[ \frac{i\hbar}{2m}\left(\Psi \boldsymbol{\nabla}\Psi^{\dagger }-\Psi^{\dagger }\boldsymbol{\nabla}\Psi \right)- \frac{q}{m}\Psi^{\dagger }\Psi \mathbf{A} + \frac{\hbar}{2m}\boldsymbol{\nabla}\times \left(  \Psi^{\dagger } \boldsymbol{\sigma } \Psi\right)\right]\cdot \delta \mathbf{A} d\tau \label{4}.
\end{eqnarray}
Identifying the integrand on the right-hand side of  Eq. (\ref{4}) with that of Eq. (\ref{1}), gives the correct expression for the Pauli current density:
\begin{equation}
\mathbf{j}=\frac{i\hbar}{2m}\left(\Psi \boldsymbol{\nabla}\Psi^{\dagger }-\Psi^{\dagger }\boldsymbol{\nabla}\Psi \right)- \frac{q}{m}\Psi^{\dagger }\Psi \mathbf{A} + \frac{\hbar}{2m}\boldsymbol{\nabla}\times \left(  \Psi^{\dagger } \boldsymbol{\sigma } \Psi\right). \nonumber
\end{equation}

\subsection{Extensions to second order}
We now consider a Hamiltonian $H(\phi,\mathbf{A})$ in the presence of an electromagnetic field
described by the scalar and vector potentials $(\phi,\mathbf{A})$:
\begin{eqnarray}
H(\phi,\mathbf{A}) &=& \frac{(\mathbf{p}-q\mathbf{A})^{2}}{2m}+q\phi -\frac{q\hbar}{2m}\boldsymbol{\sigma }\cdot \left(\boldsymbol{\nabla }\times  \mathbf{A}\right)  - \frac{q\hbar^{2}}{8m^{2}c^{2}}\boldsymbol{\nabla }\cdot\mathbf E  \nonumber \\
& & - \frac{q\hbar}{8m^{2}c^{2}}\boldsymbol{\sigma} \cdot \left[\mathbf{E}\times(\mathbf{p}-q\mathbf{A})-(\mathbf{p}  -q\mathbf{A})\times \mathbf{E}\right].
\end{eqnarray}
As before, we assume that an elementary variation of the expectation value of the energy can be associated to an elementary variation of the electromagnetic energy:
\begin{equation} \label{delta_H}
\delta \langle H(\phi,\mathbf{A}) \rangle = \int q(\rho \delta\phi  -\mathbf{j}\cdot\delta\mathbf{A})d\tau.
\end{equation}

We compute the left-hand side of Eq. (\ref{delta_H}) by evaluating separately  the variation with respect to the scalar potential $\delta_\phi H(\phi,\mathbf{A})$ and the variation with respect to the vector potential $\delta_{\mathbf{A}}H(\phi,\mathbf{A})$.
The elementary variation of the electric field can be written as: $\delta\mathbf{E}=- \boldsymbol{\nabla }\delta \phi -\partial_t \delta \mathbf{A}$, which leads to :
\begin{eqnarray}
\delta_\phi H &=& q\delta \phi +\frac{q\hbar^{2}}{8m^{2}c^{2}} \nabla\cdot \nabla\delta \phi +\frac{q\hbar}{8m^{2}c^{2}}\boldsymbol{\sigma} \cdot \left[ \nabla \delta \phi \times(\mathbf{p}-q\mathbf{A})-(\mathbf{p}  -q\mathbf{A})\times  \nabla \delta \phi \right] \label{7} \\
& & \nonumber \\
\delta_{\mathbf{A}}H&=&-\frac{q}{2m}\left(\mathbf{p}\cdot \delta \mathbf{A}+\delta \mathbf{A}\cdot \mathbf{p} \right)+ \frac{q^{2}}{m}\mathbf{A}\cdot \delta \mathbf{A}- \frac{qh}{2m}  \boldsymbol{\sigma }\cdot \left(\nabla \times \delta \mathbf{A} \right)+\frac{q^{2}\hbar}{8m^{2}c^{2}} \boldsymbol{\sigma} \cdot \left( \mathbf{E}\times  \delta \mathbf{A} -\delta \mathbf{A}\times \mathbf{E}\right)  \nonumber \\
& &   +\frac{q\hbar^2}{8m^{2}c^{2}}  \nabla\cdot \left(\partial_t \delta \mathbf{A}  \right)+\frac{q\hbar}{8m^{2}c^{2}}\boldsymbol{\sigma} \cdot \left[ \partial_t \delta \mathbf{A} \times(\mathbf{p}-q\mathbf{A})-(\mathbf{p}  -q\mathbf{A})\times \partial_t \delta \mathbf{A} \right]  \label{8}.
\end{eqnarray}

Taking the expectation value, we obtain for the variation with respect to $\phi$:
\begin{eqnarray}
\langle \delta_\phi H \rangle&=& \int \Psi^{\dagger } \left[ q\delta \phi +\frac{q\hbar^{2}}{8m^{2}c^{2}} \nabla\cdot \nabla\delta \phi +\frac{q\hbar}{8m^{2}c^{2}}\boldsymbol{\sigma} \cdot \Big( \nabla \delta \phi \times(\mathbf{p}-q\mathbf{A})-(\mathbf{p}  -q\mathbf{A})\times  \nabla \delta \phi \Big) \right]\Psi d\tau \nonumber \\
& & \nonumber \\
&=& \int q\Psi^{\dagger } \Psi \delta \phi d\tau+ \frac{q\hbar^{2}}{8m^{2}c^{2}} \int  \Psi^{\dagger} (\nabla\cdot \nabla\delta \phi) \Psi d\tau - \frac{q^2\hbar}{4m^{2}c^{2}}\int \Psi^{\dagger }
\boldsymbol{\sigma} \cdot \left(  \nabla \delta \phi\times \mathbf{A} \right)\Psi d\tau\nonumber \\
& & \nonumber \\
& &-\frac{iq\hbar^2}{8m^{2}c^{2}} \int\Psi^{\dagger }\boldsymbol{\sigma} \cdot \left[ \nabla \delta \phi \times\nabla-\nabla \times (\nabla \delta \phi) \right] \Psi d\tau \label{9}.
\end{eqnarray}

Using the identities
\begin{eqnarray} \begin{array}{lllll}
&&\int d \tau~ \Psi^{\dagger }\nabla\cdot (\nabla\delta \phi)\Psi = \int d \tau~ \nabla\cdot \nabla(\Psi^{\dagger }\Psi)\delta \phi,  \nonumber \\ \\
&&\int d \tau \Psi^{\dagger }  \boldsymbol{\sigma} \cdot \left(  \nabla \delta \phi \times \mathbf{A} \right) \Psi =  \int d \tau \left[\nabla \cdot \left(\Psi^{\dagger }\boldsymbol{\sigma}\Psi\times \mathbf{A}\right)\right] \delta \phi, \\ \\
&&\int d \tau \Psi^{\dagger } \boldsymbol{\sigma} \cdot \left[  \nabla \delta \phi \times\nabla-\nabla \times  (\nabla \delta \phi ) \right] \Psi= \int d \tau \left[ \nabla\cdot \left(\Psi^{\dagger }\boldsymbol{\sigma}\times\nabla\Psi+\nabla\Psi^{\dagger }\times\boldsymbol{\sigma}\Psi\right)\right]\delta \phi \nonumber,
\end{array}
\end{eqnarray}
one can rewrite Eq. (\ref{9}) in the following fashion
\begin{eqnarray}
\langle \delta_\phi H \rangle &=& \int q \left[ \Psi^{\dagger }\Psi +\frac{\hbar^{2}}{8m^{2}c^{2}}\Delta(\Psi^{\dagger }\Psi) - \frac{q\hbar}{4m^{2}c^{2}}\nabla \cdot \left(\Psi^{\dagger }\boldsymbol{\sigma}\Psi\times \mathbf{A}\right)\right. \nonumber \\
& & \left.   -\frac{i\hbar^2}{8m^{2}c^{2}}\nabla\cdot \left(\Psi^{\dagger }\boldsymbol{\sigma}\times\nabla\Psi+\nabla\Psi^{\dagger }\times\boldsymbol{\sigma}\Psi\right)\right]\delta \phi ~d \tau \label{10}.
\end{eqnarray}

In accordance with Eq. (\ref{delta_H}), the variation of the expectation value of the energy with $\phi$ must equal the corresponding variation in the electrostatic energy, i.e.:
\begin{equation}
\delta_\phi \langle H(\phi,\mathbf{A}) \rangle = \int q\rho \delta\phi ~d \tau.
\end{equation}
Comparing with Eq. (\ref{10}) yields the density at second order in $1/c$:
\begin{eqnarray}
\rho&=& \Psi^{\dagger }\Psi +\frac{\hbar^{2}}{8m^{2}c^{2}}\Delta (\Psi^{\dagger }\Psi) - \frac{q\hbar}{4m^{2}c^{2}}\nabla \cdot \left(\Psi^{\dagger }\boldsymbol{\sigma}\Psi\times \mathbf{A}\right) \nonumber \\
&-& \frac{i\hbar^2}{8m^{2}c^{2}}\nabla\cdot \left(\Psi^{\dagger }\boldsymbol{\sigma}\times\nabla\Psi+\nabla\Psi^{\dagger }\times\boldsymbol{\sigma}\Psi\right) \label{11},
\end{eqnarray}
which is identical to the expression that we found using the Lagrangian approach, Eq. (\ref{Eq:Prob_dens}).


Let us now turn to the current density. For the variation with respect to $\mathbf{A}$, we get from Eq. (\ref{8})
\begin{eqnarray}
\langle \delta_{\mathbf{A}} H \rangle &=&\int \Psi^{\dagger} \Big[ -\frac{q}{2m}\left(\mathbf{p}\cdot \delta \mathbf{A}+\delta \mathbf{A}\cdot \mathbf{p} \right)
+ \frac{q^{2}}{m}\mathbf{A}\cdot \delta \mathbf{A}- \frac{qh}{2m}  \boldsymbol{\sigma }\cdot \left(\nabla \times \delta \mathbf{A} \right) \nonumber\\
&+& \frac{q^{2}\hbar}{8m^{2}c^{2}} \boldsymbol{\sigma} \cdot \left( \mathbf{E}\times  \delta \mathbf{A} -\delta \mathbf{A}\times \mathbf{E}\right) +\frac{q\hbar^2}{8m^{2}c^{2}}  \boldsymbol{\nabla }\cdot \partial_t \delta \mathbf{A}  \nonumber\\
&+& \frac{q\hbar}{8m^{2}c^{2}}\boldsymbol{\sigma} \cdot \left\{ \partial_t \delta \mathbf{A}  \times(\mathbf{p}-q\mathbf{A})-(\mathbf{p}  -q\mathbf{A})\times  \partial_t \delta \mathbf{A} \right\} \Big] \Psi d\tau \nonumber\\
& & \nonumber\\
&=& -\int q\Psi^{\dagger }\left[ \frac{i\hbar}{2m}\left(\Psi \nabla\Psi^{\dagger }-\Psi^{\dagger }\nabla\Psi \right)- \frac{q}{m}\Psi^{\dagger }\Psi \mathbf{A} + \frac{\hbar}{2m}\nabla\times \left(  \Psi^{\dagger } \boldsymbol{\sigma } \Psi\right)\right]\cdot \delta \mathbf{A} d\tau  \nonumber \\
& +&\frac{q\hbar^2}{8m^{2}c^{2}}\int \Psi^{\dagger } \Psi \left(  \boldsymbol{\nabla }\cdot \partial_t \delta \mathbf{A} \right)d\tau-  \frac{iq\hbar^2}{8m^{2}c^{2}}{\int \Psi^{\dagger }  \boldsymbol{\sigma} \cdot \left( \partial_t \delta \mathbf{A}\times\boldsymbol{\nabla }-\boldsymbol{\nabla }\times \partial_t \delta \mathbf{A} \right) \Psi d \tau}  \nonumber \\
& -&\frac{q^2\hbar}{4m^{2}c^{2}}\int \Psi^{\dagger } \boldsymbol{\sigma} \cdot \left( \partial_t \delta \mathbf{A}\times\mathbf{A} \right) \Psi d \tau + \frac{q^{2}\hbar}{4m^{2}c^{2}}  \int (\Psi^{\dagger }\boldsymbol{\sigma}\Psi )\cdot \left( \mathbf{E}\times  \delta \mathbf{A} \right) d\tau. \label{12}
\end{eqnarray}

Using the identities:
\begin{eqnarray}\begin{array}{lllll}
&& \int d \tau \Psi^{\dagger } \Psi (\nabla\cdot \frac{\partial \delta \mathbf{A}}{\partial t}) = \int d \tau ~\frac{\partial}{\partial t}[\nabla(\Psi^{\dagger }\Psi)]  \cdot  \delta \mathbf{A}, \nonumber \\ \\
&& \int d \tau \Psi^{\dagger } \boldsymbol{\sigma} \cdot \left(  \frac{\partial \delta \mathbf{A}}{\partial t} \times\boldsymbol{\nabla }-\boldsymbol{\nabla }\times  \frac{\partial \delta \mathbf{A}}{\partial t}  \right) \Psi= \int d \tau \left[\frac{\partial}{\partial t}\left(\Psi^{\dagger }\boldsymbol{\sigma}\times\nabla\Psi+\nabla\Psi^{\dagger }\times\boldsymbol{\sigma}\Psi\right) \right]\cdot \delta \mathbf{A}, \\ \\
&& \int d \tau \Psi^{\dagger } \boldsymbol{\sigma} \cdot \left(   \frac{\partial \delta \mathbf{A}}{\partial t}\times\mathbf{A} \right) \Psi= \int d \tau \left[ \frac{\partial}{\partial t}\left(\Psi^{\dagger }\boldsymbol{\sigma}\Psi\times \mathbf{A} \right)\right]\cdot \delta \mathbf{A}\nonumber, \\ \\
&& \int d \tau (\Psi^{\dagger }\boldsymbol{\sigma}\Psi )\cdot \left( \mathbf{E}\times  \delta \mathbf{A} \right)= \int d \tau \left[(\Psi^{\dagger }\boldsymbol{\sigma}\Psi)\times \mathbf{E} \right]\cdot \delta \mathbf{A},\nonumber
\end{array}
\end{eqnarray}
Eq. (\ref{12}) can be written as
\begin{eqnarray}
\langle \delta_{\mathbf{A}} H \rangle &=& -\int q\Psi^{\dagger }\left[ \frac{i\hbar}{2m}\left(\Psi \nabla\Psi^{\dagger }-\Psi^{\dagger }\nabla\Psi \right)- \frac{q}{m}\Psi^{\dagger }\Psi \mathbf{A} + \frac{\hbar}{2m}\nabla\times \left(  \Psi^{\dagger } \boldsymbol{\sigma } \Psi\right) \right. \nonumber \\
& &-\frac{\hbar^2}{8m^{2}c^{2}}\frac{\partial}{\partial t}\nabla(\Psi^{\dagger }\Psi)
+  \frac{i\hbar^2}{8m^{2}c^{2}}\frac{\partial}{\partial t}\left(\Psi^{\dagger }\boldsymbol{\sigma}\times\nabla\Psi+\nabla\Psi^{\dagger }\times\boldsymbol{\sigma}\Psi\right) \nonumber \\
& & \left. +\frac{q\hbar}{4m^{2}c^{2}}\frac{\partial}{\partial t}\left(\Psi^{\dagger }\boldsymbol{\sigma}\Psi\times \mathbf{A} \right)  - \frac{q\hbar}{4m^{2}c^{2}}(\Psi^{\dagger }\boldsymbol{\sigma}\Psi)\times \mathbf{E}\right]\cdot \delta \mathbf{A} d\tau  \label{13}. \nonumber
\end{eqnarray}

The variation of the expectation value of the energy with $\mathbf{A}$ must equal the corresponding variation in the magnetic energy, i.e.:
\begin{equation}
\delta_{\mathbf{A}} \langle H(\phi,\mathbf{A}) \rangle = -q\int \mathbf{j} \cdot \delta\mathbf{A} ~d \tau.
\end{equation}
which leads to the probability current density up to second order in $1/c$:
\begin{eqnarray}
\mathbf{j}&=& \frac{i\hbar}{2m}\left(\Psi \nabla\Psi^{\dagger }-\Psi^{\dagger }\nabla\Psi \right)- \frac{q}{m}\Psi^{\dagger }\Psi \mathbf{A} + \frac{\hbar}{2m}\nabla\times \left(  \Psi^{\dagger } \boldsymbol{\sigma } \Psi\right)  - \frac{q\hbar}{4m^{2}c^{2}}(\Psi^{\dagger }\boldsymbol{\sigma}\Psi)\times \mathbf{E} \nonumber \\
& &-\frac{\hbar^2}{8m^{2}c^{2}}\frac{\partial}{\partial t}\nabla(\Psi^{\dagger }\Psi)
+  \frac{i\hbar^2}{8m^{2}c^{2}}\frac{\partial}{\partial t}\left[\Psi^{\dagger }\boldsymbol{\sigma}\times\nabla\Psi+\nabla\Psi^{\dagger }\times\boldsymbol{\sigma}\Psi\right]  +\frac{q\hbar}{4m^{2}c^{2}}\frac{\partial}{\partial t}\left(\Psi^{\dagger }\boldsymbol{\sigma}\Psi\times \mathbf{A} \right).\nonumber
\end{eqnarray}
This expression is identical to the one obtained with the Lagrangian method, Eq. (\ref{Eq:Prob_Curr}).


\begin{thebibliography}{99}

\bibitem{Bigot_annphys} J.-Y. Bigot and M. Vomir, Ann. Phys. (Berlin) {\bf 525}, 2 (2013).

\bibitem{Beaurep} E. Beaurepaire, J.-C. Merle, A. Daunois, and J.-Y. Bigot, Phys. Rev. Lett. {\bf 76}, 4250 (1996).

\bibitem{Zhang} G. P. Zhang and W. H\"{u}bner, Phys. Rev. Lett. {\bf 85}, 3025 (2000).

\bibitem{Koopmans} B. Koopmans, J. J. M. Ruigrok, F. Dalla Longa, and W. J. M. de Jonge, Phys. Rev. Lett. {\bf 95}, 267207 (2005).

\bibitem{Bigot_natphys} J.-Y. Bigot, M. Vomir, and E. Beaurepaire, Nature Phys. {\bf 5}, 515 (2009).

\bibitem{Vonesch} H. Vonesch and J.-Y. Bigot, Phys. Rev. B {\bf 85}, 180407 (2012).

\bibitem{Hinsch_prb} Y. Hinschberger and P.-A. Hervieux, "Classical
Modeling of Ultrafast Coherent Magneto-Optical Experiments",
submitted to Phys. Rev. B (2013).

\bibitem{Keller} S. Keller, Colm T. Whelan, H. Ast, H. R. J. Walters, and R. M. Dreizler, Phys. Rev. A {\bf 50}, 3865 (1994).

\bibitem{Dyall} Kenneth G. Dyall and Knut F{\ae}gri, Jr., {\it Introduction to relativistic quantum chemistry} (Oxford University Press, Oxford, 2007).

\bibitem{Engel} E. Engel, R. M. Dreizler, S. Varga, and B. Fricke, in {\it Relativistic Effects in Heavy-Element Chemistry and Physics}, ed. B. A. Hess (Wiley, New York, 2003) p. 123--161.

\bibitem{Rajagopal} A. I. Rajagopal, J. Phys. C: Solid State Phys. {\bf 11}, L943 (1978).

\bibitem{Parpia} F. A. Parpia and W. R. Johnson, J. Phys. B: At. Mol. Phys. {\bf 17}, 531 (1984).

\bibitem{Romaniello} P. Romaniello and P. L. de Boeij, J. Chem. Phys. {\bf 127}, 174111 (2007).

\bibitem{Foldy} L. L. Foldy, Phys. Rev. {\bf 87}, 688 (1952).

\bibitem{Hinsch} Y. Hinschberger and P.-A. Hervieux, Phys. Lett. A {\bf 376}, 813 (2012).

\bibitem{Van_Lenthe} E. van Lenthe, E. J. Baerends, and J. G. Snijders, J. Chem. Phys. {\bf 99}, 4597 (1993).

\bibitem{Holland-Brown} P. Holland and H. R. Brown, Studies in History and Philosophy of Modern Physics {\bf 34}, 161 (2003).

\bibitem{Sulaksono} A. Sulaksono, P.-G. Reinhard , T. J. B\"{u}rvenich, P. O. Hess, and  J. A. Maruhn,  Phys. Rev. Lett. {\bf 98}, 262501 (2007).

\bibitem{manfredi_maxwell} G. Manfredi, Eur. J. Phys. {\bf 34}, 859 (2013).

\bibitem{Durr} H. P. D\"{u}rr, Nuovo Cimento {\bf 22}, 386 (1974).

\bibitem{Landau}
    L. D. Landau and E. M. Lifshitz, {\it Quantum Mechanics: Non-Relativistic Theory} (Butterworth-Heinemann, Oxford, 1977).

\bibitem{Zamanian} F. A. Asenjo, J. Zamanian, M. Marklund, G. Brodin, and P. Johansson, New J. Phys. {\bf 14} 073042 (2012).

\bibitem{Strange} P. Strange, {\it Relativistic Quantum Mechanics} (Cambridge University Press, Cambridge, 2005).

\bibitem{Sakurai} J. J. Sakurai, {\it Advanced Quantum Mechanics}, (Addison-Wesley, Reading, 1967).

\bibitem{LBLL} M. Le Bellac M and J.-M. L\'{e}vy-Leblond,  Nuovo Cimento B {\bf 14}, 217 (1973).

\bibitem{Rousseaux} G. Rousseaux,  Europhys. Lett. {\bf 71} 15 (2005)

\bibitem{Montigny2006} M. de Montigny and G. Rousseaux, Eur. J. Phys. {\bf 27} 755 (2006).

\bibitem{Kiefer} D. Kiefer et al., Nature Comm. {\bf 4}, 1763 (2013).

\bibitem{Haines} M. G. Haines, M. S. Wei, F. N. Beg, and R. B. Stephens, Phys. Rev. Lett. {\bf 102}, 045008 (2009).

\bibitem{Ridgers} C. P. Ridgers, C. S. Brady, R. Duclous, J. G. Kirk, K. Bennett, T. D. Arber, and A. R. Bell, Phys. Plasmas {\bf 20}, 056701 (2013).


\end{thebibliography}
\end{document}